\documentclass[preprint,review,12pt]{elsarticle}

\usepackage{graphicx, subcaption}
\usepackage{epstopdf}
\usepackage{dcolumn}
\usepackage{color}
\usepackage{multirow}
\usepackage{textcomp,gensymb}
\usepackage{physics}
\usepackage{epsfig}
\usepackage{amsmath}
\usepackage{amssymb}
\usepackage{booktabs}
\usepackage{siunitx}
\usepackage{breqn}
   \usepackage[utf8]{inputenc}
    \usepackage{tabularray}
    \usepackage[table, xcdraw]{xcolor}
\usepackage{ragged2e}                       % new
\usepackage{booktabs, 
            makecell, multirow, tabularx} 
\usepackage{ragged2e}
\usepackage[percent]{overpic}
\usepackage{hyperref}
\usepackage{xr}
\biboptions{square,comma,sort&compress}
\journal{}
\begin{document}
\begin{frontmatter}
%\title{Magnetic Field-Assisted Grain Boundary Migration in Polycrystalline Non-magnetic Thin Films: A Phase-Field study}
%\title{Grain Boundary Grooving in Polycrystalline Non-magnetic Thin Films: Effect of Magnetic Field}
%\title{Universal Behavior of Surface Grooving in Thin Non-magnetic Films Under External Magnetic Field: Insights from Phase-Field approach}

%\title{Thermal Grooving in Thin Non-magnetic Films: Unraveling the Universal Nature Under External Magnetic Field}
\title{\textcolor{black}{Grain boundary grooving in thin film under the influence of an external
magnetic field: A phase-field study}}
%\title{Magnetic Field-Assisted Grain Boundary Migration in the presence of thermal grooving in a non-magnetic system: A phase-field study}

\author[a,b]{Soumya Bandyopadhyay}
\author[b]{Somnath Bhowmick$^\ast$\corref{cor}}
\ead{bsomnath@iitk.ac.in}
\author[b]{Rajdip Mukherjee$^\ast$\corref{cor}}
\ead{rajdipm@iitk.ac.in}
\address[a]{Department of Materials Science and Engineering, University of Florida, Gainesville, Florida-32611, United States}

\address[b]{Department of Materials Science and Engineering, Indian Institute of
Technology, Kanpur, Kanpur-208016, UP, India} 

\begin{abstract}
Using a phase-field model, we study the surface diffusion-controlled grooving of a moving grain boundary under the influence of an external magnetic field in thin films of a non-magnetic material. The driving force for the grain boundary motion comes from the anisotropic magnetic susceptibility of the material, leading to the free energy difference between differently oriented grains. We find that, above a critical magnetic field, the grain boundary motion is in a steady state, and under this condition, the mobile thermal groove exhibits a universal behavior - scaled surface profiles are time-invariant and independent of thermodynamic parameters. 
The simulated universal curve agrees well with Mullins’ theory of mobile grooves for any groove shape. We extend our study to a three-dimensional polycrystalline thin film with equal-sized hexagonal grains. We observe a preferential grain growth depending on the applied magnetic field direction, which can be leveraged for field-assisted texture control of polycrystalline thin films. 
Our study reveals that keeping other conditions the same, the rate of pitting at the vertices of the hexagonal grains substantially decreases in the presence of the external magnetic field.
\end{abstract}

%%Graphical abstract
%\begin{graphicalabstract}
%\includegraphics[width=1.2\linewidth]{Graphical_abs.pdf}
%\end{graphicalabstract}

%%Research highlights
%\begin{highlights}
%\item Our study displays a preferential grain growth depending on the applied magnetic field direction and the universal behavior of a mobile thermal groove when grain boundary motion is in a steady state above a critical magnetic field. It also reveals the different regimes of grain boundary motion at different magnetic fields. During the steady-state motion of the GBs, the normalized asymmetric surface profiles show a universal behavior and exhibit excellent quantitative agreement with Mullins’ theory of mobile grooves.

%\item Our studies on 3D polycrystalline systems with equal-sized hexagonal grains dictate a way to control the growth of differently oriented grains by controlling the direction of the applied magnetic field. We also show the pitting phenomenon and the intricate relationship and interaction between the pits and the grooves formed in a thin polycrystalline film with a free surface
%\end{highlights}

\end{frontmatter}

\section{\label{sec:intro} Introduction}
Grain boundaries (GBs), potential sites for morphological instabilities, significantly influence the microstructural properties in metallic systems. Their inherent susceptibility to external fields such as thermal, electrical, 
magnetic, and mechanical forces renders them prone to morphological instabilities. These boundaries also play a pivotal role in facilitating various phenomena, including the formation of void and hillock, grain boundary migration, and acting as sites for heterogeneous nucleation of secondary phases, etc.~\cite{huntington1961current,black1969electromigration, winning2001stress,winning2002mechanisms,tonks2010analysis,bhattacharyya2011phase}.

Furthermore, electric or thermal field-induced migration of GBs, or a combination of both, poses an enduring challenge for the long-term reliability of microelectronic packaging and integrated circuit (IC) industries. This reliability issue is intricately associated with the formation of voids or the emergence of hillocks, emphasizing the critical need for robust and optimized solutions to ensure improved performance and durability~\cite{ballo2001grain, hadian2018migration, chen2020atomistic, liang2021effect, ho1989electromigration, chakraborty2018phase, tu1992electromigration, mukherjee2016phase, hadian2018migration}. For example, in Al-Cu flip-chip solder joints, electromigration associated with thermomigration can manifest hillock and void formation, leading to short and open failures in the systems~\cite{chen2010electromigration,hu2003electromigration}. In some scenarios, the presence of stress along with thermal and electrical load can also cause the materials to fail~\cite{lee1999analysis, gungor1998modeling, sarychev1999general}.

Recent studies report that using an external magnetic field provides additional parametric quantity to tune the grain structure and the material properties in various materials~\cite{backofen2019controlling,molodov2014grain}. Even though the complex synergism between 
the atomic diffusion, irreversible deformation mechanisms, and magnetic influence have been reported (see the comprehensive review by Guillon \emph{et al.} and the reference therein~\cite{guillon2018manipulation}), gaining a thorough comprehension of the interplay between magnetic fields and the transport of solid-state matter remains a challenging task that demands further exploration.~\cite{backofen2019controlling}. 

The initial observation of magnetic field effects on GB migration is credited to Mullins.~\cite{mullins1956magnetically}. In ferromagnetic materials, magnetic effect on grain growth is more substantial due to the presence of magnetocrystalline anisotropy~\cite{backofen2019controlling}. When subjected to a strong external magnetic field, magnetic materials align their inherent magnetic moments accordingly. Materials with anisotropic magnetic properties exhibit variations in bulk free energy among differently oriented grains, potentially instigating grain boundary migration.~\cite{backofen2019controlling}.

%In some scenarios, stress-assisted diffusion of vacancies can lead to cavity formation (creep cavitation)~\cite{perry1974cavitation,kassner2003creep} or stress-assisted instabilities (Asaro–Tiller–Grinfeld (ATG) instabilities)~\cite{yeon2006phase,chirranjeevi2009phase} in thin films.

%Grain boundaries are potential sites for morphological instabilities and are highly vulnerable to external thermo-electro-magneto-mechanical load. Grain boundary migration (electromigration and thermomigration), leading to void or hillock formation, has always been long-term reliability issues in the microelectronic packaging and integrated circuit (IC) industries~\cite{ho1989electromigration, chakraborty2018phase, tu1992electromigration, mukherjee2016phase}. For example, electromigration and thermomigration can lead to failures in Al-Cu flip-chip solder joints~\cite{chen2010electromigration}. \textcolor{red}{Stress-induced instabilities (Asaro–Tiller–Grinfeld (ATG) instabilities)~\cite{yeon2006phase,chirranjeevi2009phase} and stress-assisted diffusion of vacancies leading to cavity formation (creep cavitation)~\cite{perry1974cavitation,kassner2003creep} and failure, are also common in thin films.} 

More interestingly, some findings demonstrate magnetic field-induced migration of the GBs, even in diamagnetic systems such as Bi, Zn, Sn, Ti, etc, possessing anisotropic crystal structure~\cite{mullins1956magnetically, molodov1997motion,sheikh2003migration,molodov2004grain, molodov2014grain}. For instance, research indicates that the orientation of Sn within Sn-Ag-Cu solder interconnects can be influenced by magnetic fields,  potentially leading to significant implications for the material's performance and reliability~\cite{chen2015magnetic}. Another study by He \emph{et al.}~\cite{he2015evolution} demonstrates the impact of a strong magnetic field on copper. Their findings
suggest that employing an external magnetic field is a viable approach for controlling the evolution of polycrystalline microstructures in non-ferromagnetic materials.
 Theoretical studies using the multiphase field model (MPF) by Rezaei \emph{et al.}~\cite{rezaei2021phase} reported the microstructure evolution of titanium polycrystalline system under an external magnetic field. They adopted a columnar microstructure to perform the grain growth simulations in bicrystalline as well as polycrystalline Titanium systems, corroborating their results with Molodov \emph{et al.}'s experimental observations~\cite{molodov1998true,sheikh2003migration}.

Although most of the studies mainly focus on the influence of magnetic fields on bulk systems, no proper investigation reports the case of polycrystalline thin films with free surfaces subjected to an applied external magnetic field. The primary mechanism responsible for the rupture of polycrystalline thin films entails the formation, deepening, and complex interplay of grooves and pits at the multi-junctions~\cite{gonzalez2011morphology, thompson1990grain, frost1994microstructural, lin2016modified,mukherjee2022grain}. When subjected to high temperatures, polycrystalline thin films even exhibit the formation of grain boundary grooves at the free surface. 
 Nonetheless, dynamics of thermal groove formation are intricately governed by evaporation condensation as well as atomic diffusion at the surfaces, grain boundaries, and lattices~\cite{mukherjee2022grain}. 
 \textcolor{black}{In his classical paper, W. W. Mullins described the formation of surface grooves at the grain boundaries of a heated polycrystal as ``thermal grooving"~\cite{mullins1957theory}. In this study, Mullins pioneered the concept of the formation of thermal grooves, occurring at the confluence of stationary GB and the film surface at elevated temperatures.}
His findings suggest the groove retains a form irrespective of time ($t$), with deepening rates scaling as $t^{\frac{1}{4}}$ 
when surface diffusion dominates solute transfer, and $t^{\frac{1}{2}}$ when the evaporation-condensation 
governs the solute transfer mechanism~\cite{mullins1957theory,mullins1958effect}.

Mullins further developed his analysis to address the behavior of moving grain boundaries, outlining three 
possible scenarios: (a) complete immobilization of the boundary by the groove, (b) steady-state motion where 
the groove progresses alongside the GB at a uniform velocity, and (c) decelerating motion characterized by a 
gradual decrease in the grain boundary velocity over time due to partial stagnation caused by the groove.
Additionally, Mullins distinguishes the asymmetrical form of the boundary's steady-state motion, in contrast to the complete stagnation with the similar surface profile observed for stationary scenerio~\cite {mullins1958effect}. The analytical 
definition of the characteristic shape as a function of GB energy $\gamma_{GB}$, surface energy $\gamma_s$, 
and steady-state grain boundary velocity can be obtained in the 
classic work by Mullins~\cite{mullins1958effect}. Experimental~\cite{munoz2004monitoring}, and analytical 
investigations on polycrystalline ceramics and metallic films report such dynamic groove profile asymmetry due to anisotropic surface energy~\cite{min2006effect}.
Furthermore, asymmetric surface flux may result in asymmetric surface profiles owing to various factors, including applied electric, thermal, magnetic, and convective flow, etc.~\cite{verma2022computational}. However, the factors manifesting in the groove's universality and its time-invariant form in these scenarios require critical exploration. Due to their universal behavior, stationary thermal grooves are used to quantify the surface diffusion coefficients of mobile species as a function of temperature, which are compared with Mullin's analytical model. These universal profiles can further be used to calculate the ratio between GB energy $\gamma_{GB}$ and surface energy $\gamma_{s}$.

Recently, Verma \emph{et al.} reported on the universal nature of the surface profiles of a mobile groove, demonstrating its correlation with surface diffusivity, GB mobility, and the thickness of the film~\cite{verma2022computational, verma2023effect}. According to their findings, during steady-state 
grain boundary motion, universal behavior persists regardless of film thickness, surface diffusivity, and 
grain boundary mobility. Additionally, their results substantiate Mullins' theory of dynamic grooves across different groove forms.
However, their study uses some hypothetical assumptions regarding the free energy difference 
between the grains, which triggers the GB migration. 

%However, one should note that previous explorations were limited to studying the universal behavior of a stationary thermal groove~\cite{verma2022computational, verma2023effect}.
%by assuming a free energy difference between the grains 
%However, their study triggered the GB migration by assuming a free energy difference between the grains, thus altering their thermodynamic stability. Although they did not study the effect of the external field on GB migration, in their analytical definition, they mentioned the effect of the external field. 

In this study, we use an external magnetic field to trigger the GB migration, which is doable in reality. We find that an external magnetic field provides physical control of the grain boundary migration and the nature of the corresponding groove profiles. We perform two-dimensional (2D) phase field simulations considering a bi-crystalline diamagnetic film in equilibrium with its vapor phase. Furthermore, we extend our model to three dimensions (3D) to simulate pitting at the GB triple junctions in a polycrystal diamagnetic film. In the next section, we derive the formulation required for the phase field simulations. In the subsequent sections, we discuss our significant findings, and later, we summarize the key conclusions from our study.
%instead of a hypothetical assumption of the alteration of the free energies.

\section{\label{sec:formula} Formulation}
The phase-field model has emerged as a powerful and reliable computational tool for modeling and predicting 
mesoscopic morphological evolution based on the thermodynamics and kinetics of materials for the last few 
decades~\cite{chen2002phase}. This approach can also predict the microstructure as well as the evolution under 
external stimuli without a priori assumptions on the system 
morphologies~\cite{chen2002phase,krill2002computer,mukherjee2022grain, 
chafle2022domain,PhysRevMaterials.7.083802,bandyopadhyay2023role,bandyopadhyay2023multivariant}. In general, it 
describes a microstructure using a set of conserved and nonconserved field variables or order parameters, which 
are continuous across the interface~\cite{chen2002phase}. Thus, the phase-field model becomes a perfect choice 
for studying grain boundary grooving in the realm of the mesoscopic scale~\cite{mukherjee2022grain}. Recently, there have been numerous reports utilizing phase-field models to investigate complex grain boundary grooving 
problems~\cite{bouville2006phase,bouville2007grain,moelans2007phase,joshi2017destabilisation,farmer2020phase}.

This section briefly describes the model employed for 2D and 3D simulations. Previously, Mukherjee et al. utilized this model for the investigation of sintering kinetics of nanoparticle aggregates~\cite{mukherjee2011thermal}. Later, Mukherjee et al. extended this model to study grain boundary grooving~\cite{mukherjee2016phase, mukherjee2022grain}. The model employs density, $\rho(\mathbf{r})$  as a conserved field variable to distinguish between the film ($\rho(\mathbf{r}) = 1$) and the vapor phase ($\rho(\mathbf{r}) = 0$). Additionally, we use multiple non-conserved order parameters $\eta_{i}(\mathbf{r})$; $(i = 1, 2,\dots N)$ to represent $N$ grains characterized by distinct orientations. Here $\mathbf{r}$ defines the position vector. Hence, we define the total free energy of the system as~\cite{mukherjee2022grain}:
\begin{equation}
    \mathcal{F} = \int_{V}\left[f_{0}(\rho,\eta_1, \eta_2,\ldots \eta_N) + \kappa_{\rho}(\mathbf{\nabla}\rho)^2 + \kappa_{\eta}\sum_{i=1}^{N}(\mathbf{\nabla}\eta_{i})^2 + f_{m}\right]dV,
    \label{eq1}
\end{equation}
where $f_{0}(\rho,\eta_1, \eta_2,\ldots \eta_N)$ is the bulk free energy of the system. The second and the third terms relate the gradient energy contributions in $\rho(\mathbf{r})$ and $\eta_{i}(\mathbf{r})$, respectively. $\kappa_{\rho}$ and $\kappa_{\eta}$ define the gradient energy coefficients of the corresponding order parameters ($\rho(\mathbf{r})$, $\eta_{i}(\mathbf{r})$). The fourth term $f_m$ in Eq.~\eqref{eq1} denotes the magnetic contribution to the free energy associated with the magnetic migration.

The bulk free energy $f_{0}(\rho,\eta_{i})$ is defined as:
\begin{equation}
    f_{0}(\rho(\mathbf{r}),\eta_{i}(\mathbf{r}))= A\rho^2(1-\rho)^2 + B\rho^2\xi(\eta_{i}) + C(1-\rho)^2\sum_{i=1}^{N}\eta_{i}^2,
\label{eq2}
\end{equation}
where
\begin{equation}
    \xi(\eta_{i}) = \sum_{i=1}^{N}\left(\frac{\eta_{i}^4}{4} - \frac{\eta_{i}^2}{2} + \Gamma\sum_{j > i}\eta_i^2\eta_j^2 + 0.25\right),
\label{eq3}
\end{equation}
and $A$, $B$, $C$, $\Gamma$ are the free energy constants. We construct the bulk free energy $f_{0}(\rho(\mathbf{r}),\eta_{i}(\mathbf{r}))$,  to ensure equilibrium between the solid and vapor phases across surfaces and GBs.
 This form of energy density is one of the simplest possible forms manifesting equilibrium properties by just tuning the interfacial energy or width with an appropriate choice of free energy constants~\cite{mukherjee2022grain}.

Additionally, incorporating contribution from grains with different orientations, we introduce the magnetic energy contribution $f_m$ as~\cite{liang2022phase}:
\begin{equation}
   f_m = \sum_{i=1}^{N}h(\eta_i)f_{m,i}.
\label{eq4}
\end{equation}
Here $h(\eta_{i}) = \eta_{i}^3(6\eta_{i}^2 - 15\eta_{i} + 10)$ is the switching function associated with the order parameters (grains). Moreover, we express the magnetic part of grain $i$ as $f_{m,i} = \mu_{0}\textcolor{red}{\mathbf{H}}^2\chi^{i}/2$. Here $\mathbf{H}$ defines the applied magnetic field. $\chi^{i}$ and $\mu_{0}$ correspond to the susceptibility (magnetic) and permeability (magnetic) of grain $i$. Note that in anisotropic systems, the second-rank tensor $\chi^{i}$ relies on the grain orientations, and we have used boldfaced letters to define the vector quantities.

\begin{figure*}[htbp]
    \subfloat[2D Cartesian frame of reference]{\label{fig:2dmag}\includegraphics[width=.43\linewidth]{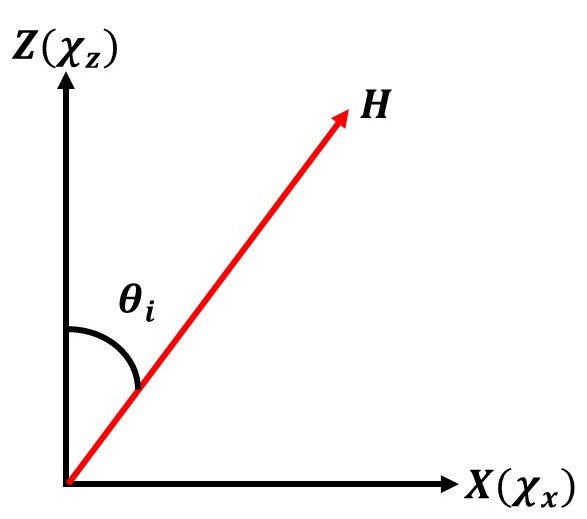}}\hfill
\subfloat[3D Cartesian frame of reference]{\label{fig:3dmag}\includegraphics[width=.45\linewidth]{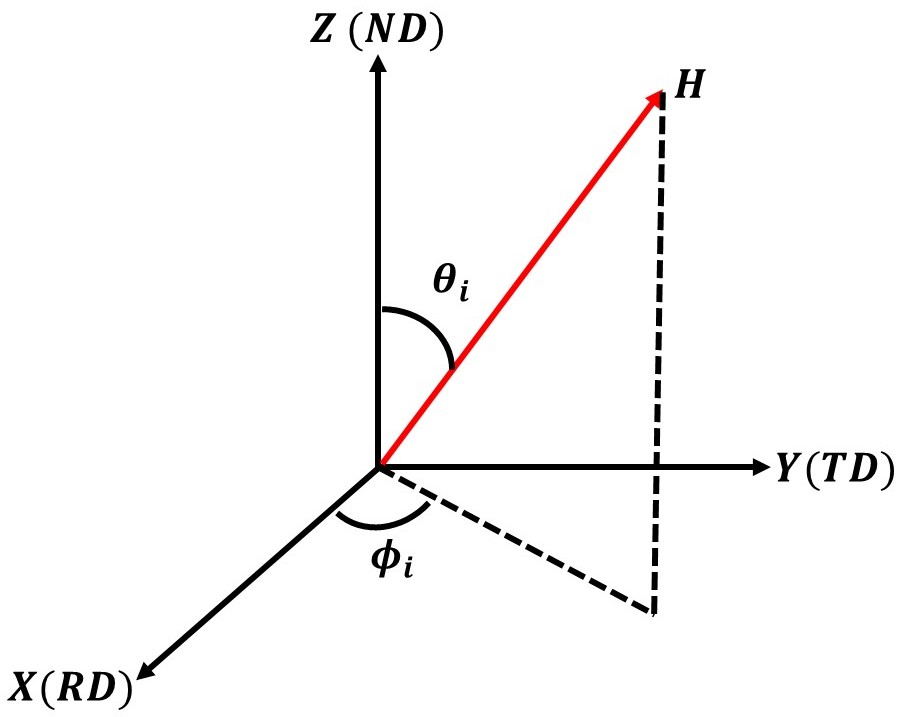}}\hfill 
    \caption{\textcolor{black}{Schematic depiction of the \textbf{H} vector in both two-dimensional (2D) and three-dimensional (3D) settings. (a) Shows the component form of \textbf{H} in 2D. Here, $\chi$ denotes susceptibility, and $\theta_i$ represents the angular relationship between the Z-axis associated with each grain and $\textbf{H}$. Where $i = 1, 2, \ldots, N$ represents the number of distinct grain orientations. (b) Depicts the 3D scenario where the magnetic field vector $\mathbf{H}$ forms angles $\theta$ with respect to the ND (Z) axis and $\phi$ with the RD (X) axis, projecting onto the RD (X)-TD (Y) plane.}}
    \label{fig:dimenmag}
\end{figure*}

The present study focuses on $\textrm{Bi}$, a diamagnetic material possessing hexagonal crystal structure; therefore we can simplify the susceptibility as $\chi_{a}^{i} = \chi_{b}^{i}$, $\chi_{c}^{i}$. Thus, as suggested by Liang et al.~\cite{liang2022phase}, in a two-dimensional (2D) scenario we use the component form of the magnetic field considering $\chi_{z}^{i}$, $\chi_{x}^{i}$ aligned with the $Z$ and $X$ axis (shown as a schematic in Fig.~\ref{fig:2dmag}), respectively:
\begin{equation}
   \mathbf{H} = H_{m}sin\theta_{i}\hat{n_x} + H_{m}cos\theta_{i}\hat{n_z},
\label{eq5}
\end{equation}
where $\theta_{i}$ signifies the angular relationship between the $Z$-axis associated with the specific grain and the orientation of the magnetic field. In our simulations, we align the crystallographic $a$ (or $c$) axis 
and $c$ (or $a$) axis along the $X$ and $Z$ directions, respectively. $\hat{n_x}$, $\hat{n_z}$ denotes the unit 
vectors along the $X$ and $Z$ directions, and the magnetic field amplitude is given by $H_m$. Therefore, we rewrite the 
magnetic contribution as~\cite{liang2022phase}:
\begin{equation}
  f_{m,i} = \mu_{0}H_m^2(\chi_{x} - cos^2\theta_{i}(\chi_z^{i} - \chi_{x}^{i}))/2
\label{eq6}
\end{equation}

%\begin{figure}[!htpb]
%   \centering
%    \includegraphics[scale = 0.3]{schematic2D.jpg}
%   \caption{Schematic representation of the $\mathbf{H}$ vector in the 2D Cartesian frame of reference.
%     }
%  \label{fig:2dmag}
%\end{figure}

However, in real scenarios, for example, considering metal sheet samples or in three-dimensional (3D) cases, one can also consider the direction of the applied magnetic field with respect to the rolling (RD), transverse (TD), and normal (ND) directions~\cite{rezaei2021phase}. Note that here, we use the terms RD, TD, and ND as a general description of the Cartesian coordinates (X, Y, Z), and we do not consider any textured system in this study. Let us define the direction of the magnetic field $\mathbf{H}$ in such a way that it exhibits an angle $\theta$ with respect to the ND (Z) and $\phi$ with RD (X) (projection of the magnetic field direction on the RD (X)-TD (Y) plane). Thus, the magnetic field vector $\mathbf{H}$ can be defined as (shown as schematic in Fig.~\ref{fig:3dmag}):
   \begin{equation}
   \mathbf{H} = H_{m}sin\theta_{i} cos\phi_{i}\hat{n_x} + H_{m}sin\theta_{i} 
   sin\phi_{i}\hat{n_y} + H_{m}cos\theta_{i}\hat{n_z},
\label{h3d}
\end{equation}
where $\hat{n_x}$, $\hat{n_y}$, $\hat{n_z}$ correspond to the unit vectors along the $X$, $Y$ and $Z$ directions, respectively. Finally, in this case, we rewrite the form of the magnetic contribution as:
\begin{equation}
  f_{m,i} = \frac{\mu_0 H_m^2}{2}\left[\left\{\chi_y^i + (\chi_x^i - \chi_y^i)cos^2\phi_{i}\right\} + cos^2\theta_{i}\left\{\chi_z^i - (\chi_y^i + (\chi_x^i - \chi_y^i)cos^2\phi_{i})\right\}\right].
\label{magE3d}
\end{equation}
\textcolor{black}{Here $\chi_{x}^{i}$, $\chi_{y}^{i}$, and $\chi_{z}^{i}$ represent the susceptibility along X, Y, and Z axes.}

Having obtained all the required free energies to study the spatiotemporal evolution of the conserved and nonconserved order parameters, we solve the coupled Cahn–Hilliard (CH) and Allen–Cahn (AC) equations using a 2D/3D simulation grid~\cite{mukherjee2016phase, mukherjee2022grain, verma2022computational,verma2023effect}:
\begin{align}
 &\frac{\partial \rho}{\partial t} = \mathbf{\nabla}\cdot (M\nabla \mu),\\
 &  \frac{\partial \eta_{i}}{\partial t} = -L\frac{\delta \mathcal{F}}{\delta \eta_{i}}.
\label{eq7}
\end{align}
$M$ defines the mobility of the chemical species, $\mu$ is the generalized chemical potential, and $L$ defines the relaxation coefficient related to order parameter field. Thus, we rewrite the chemical potential as follows:
\begin{equation}
 \mu =\frac{\delta \mathcal{F}}{\delta \rho} = \frac{\partial f_{0}}{\partial \rho} - \kappa_{\rho}\mathbf{\nabla}^2 \rho.
\label{eq8}
\end{equation}
In this study, we consider the atomic mobility at the surface by considering it as a function of the density field as~\cite{verma2021grain}:
\begin{equation}
    M (\rho) = M_{b} + 16M_{s}\rho^2(1-\rho)^2.
    \label{eq9}
\end{equation}
Here $M_{b}$ denotes the bulk mobility, and $M_{s}$ corresponds to the surface mobility related to the surface diffusivity. We take $M_s$ to be $10^3$ to $10^6$ higher than the $M_b$. We choose $L/M_s$ such that we are always in the regime where the diffusion mechanism is surface diffusion dominated~\cite{verma2022computational, mukherjee2022grain}. The asymptotic interpretation of phase field equations (CH and AC) are performed by Mukherjee et al. by considering the motion by surface Laplacian of mean curvature alongside AC dynamics~\cite{mukherjee2019electric}. Their study shows %$B$ corresponds to $\gamma_s M_s$ and 
for a particular value of $M_s$, $L$ must ensure the surface diffusion-controlled motion, i.e., groove depth $dg$ remains proportional to $t^{1/4}$~\cite{mukherjee2019electric,verma2022computational, mukherjee2023phase}. We solve all the equations in their corresponding weak forms using an open-source finite element `MOOSE'  framework~\cite{schwen2017rapid,permann2020moose}, assuming periodic boundary conditions in all directions. For more details regarding the weak form and implementation in MOOSE, please refer to~\cite{lindsay2022moose,schwen2017rapid,permann2020moose}. All the parameters used in the simulations are tabulated in ~\ref{param} of this article. 

\section{\label{sec:res} Results and Discussions}
%In this section, we discuss our findings. We present our results in two steps. First, we discuss the effect of an applied magnetic field on a two-dimensional bicrystal and try to validate our findings with the analytical reports obtained by Mullins. In the next step, we extend our model to study a three-dimensional polycrystalline hexagonal thin film subjected to the applied magnetic field. Since a three-dimensional polycrystalline thin film with a free surface is prone to create grooves (at the grain boundaries) as well as pits (at the junctions), we also try to study the complex interplay of the groove-pit behavior under external magnetic load.
\begin{figure*}[htbp]
\subfloat[time = 0 ]{\label{zoomed}\includegraphics[width=0.8\linewidth]{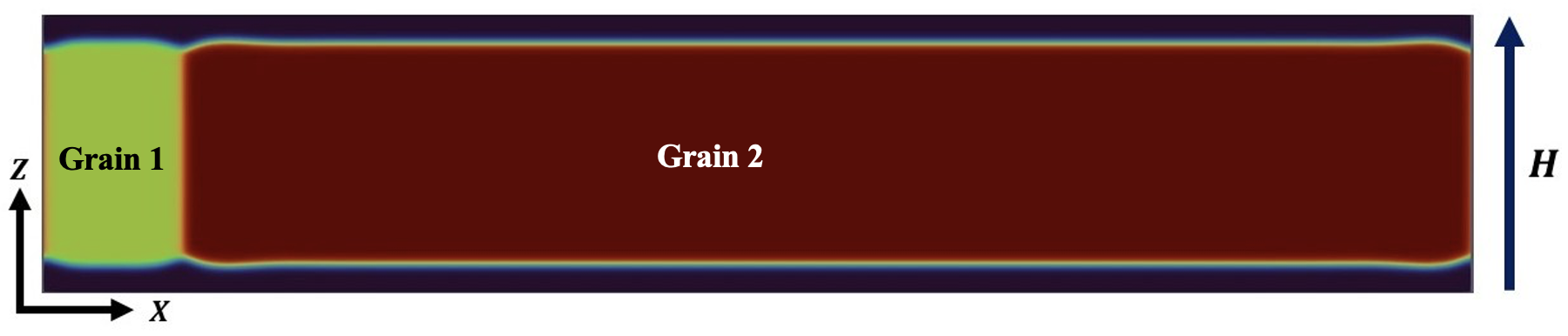}}\hfill
\subfloat[time = 2000]{\label{free2}\includegraphics[width=.74\linewidth]{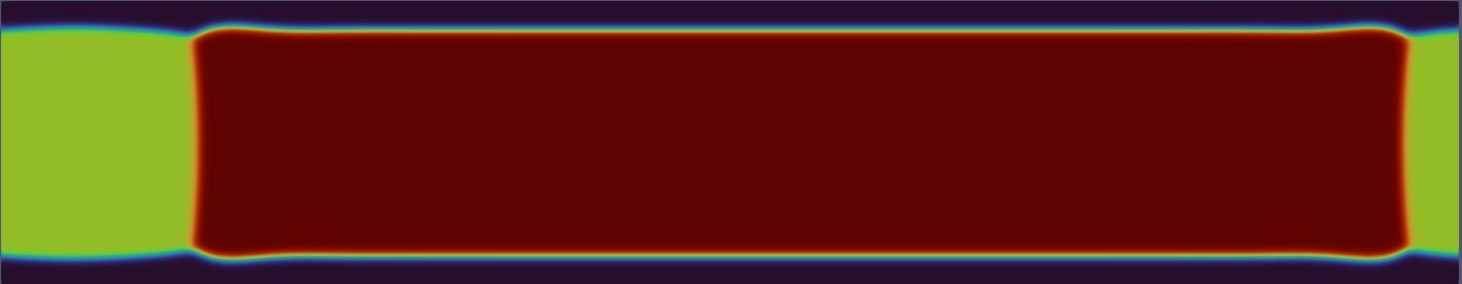}}\hfill
\subfloat[time = 10000]{\label{free3}\includegraphics[width=.74\linewidth]{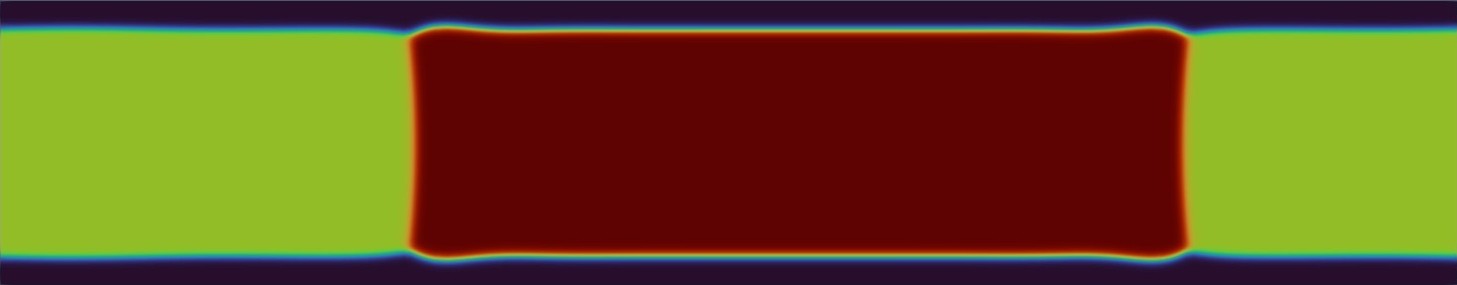}}
\caption{Dynamic microstructural evolution showing the migration of GB within a bicrystal over time. The direction of the magnetic field is parallel ($\mathbf{H\parallel}$) to the $Z$ axis.}
\label{fig:free_comp}
\end{figure*}
\subsection{2D simulations of grain boundary grooving under applied magnetic field }
\label{res2d}
We begin with a two-dimensional bicrystal morphology, as shown in Fig.~\ref{zoomed}. 
\textcolor{black}{Note that, while we perform all the simulations considering an isothermal scenario, the temperature is chosen such that atomic migration primarily occurs through surface diffusion}.
Since we mainly aim to study the magnetic synergism with the groove morphology, we consider a system with a preformed groove (Fig.~\ref{zoomed}) at the confluence of the grain boundaries and free surface. 
Moreover, considering such morphology provides an adequate temporal allowance to study a proper magnetic field-induced grain boundary-assisted groove migration. We consider a system size of $1024\Delta x \times 120\Delta z$, consisting of a free-standing film of thickness $80\Delta z$ and a vapor layer of thickness $20\Delta z$ on the top and bottom of the film surface, where $\Delta x$, $\Delta y$, $\Delta z$ are the grid spacings in the $X$, $Y$, and $Z$ directions respectively. We employ two grains oriented differently: Grain 1 at $0^\circ$ (colored in lemon green) and Grain 2 at $90^\circ$ (colored in brown), aiming to maximize anisotropic contribution in the magnetic part (as illustrated in Fig.~\ref{zoomed}). It is crucial to emphasize that the grain oriented at $0^\circ$ features its c-axis aligned parallel to the Z-direction of the Cartesian frame of reference. Conversely, in the case of the grain-oriented at $90^\circ$, its c-axis aligns perpendicular to the Z-direction or, in the context of 2D simulations, parallel to the X-direction.
%\begin{figure}[!htpb]
%   \centering
%    \includegraphics[scale = 0.3]{t01.png}
%   \caption{Bicrystal morphology with a preformed groove considered as the initial condition for the subsequent simulations.
%     }
%  \label{fig:init2d}
%\end{figure}

\begin{figure*}[!htpb]
\subfloat[]{\label{fig:grv3a}\includegraphics[width=0.6\linewidth]{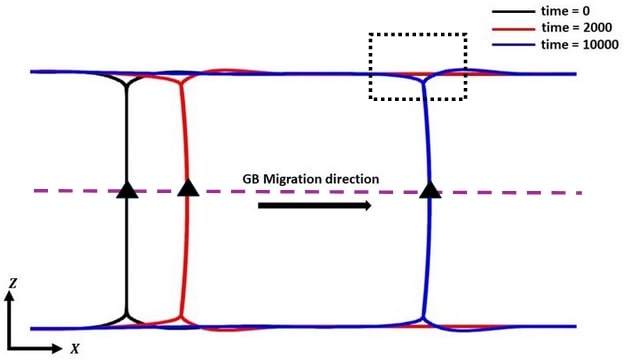}}\hfill
\subfloat[]{\label{fig:grv3b}\includegraphics[width=0.4\linewidth]{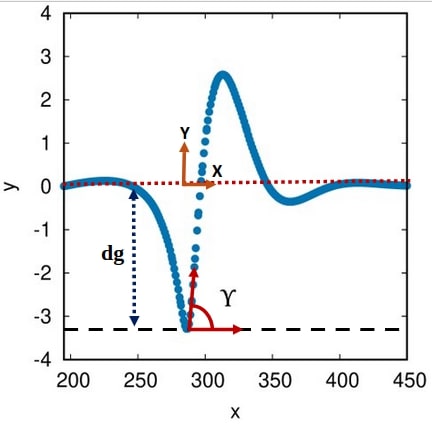}}\hfill 
\caption{Temporal evolution of a mobile groove attached to the GB. The groove creates an asymmetric profile as it migrates toward the right under external magnetic field $H_{m} = 4.0\times 10^8 A/m$.
\textcolor{black}{(a) Grain boundary (GB) migration under magnetic field. The dashed line represents the mid-plane of the film and the points are marked to calculate the GB position. The dotted rectangular box displays the representative element used to perform analysis shown next. (b) Schematic representation of the grain boundary groove for a steady-state scenario (zoomed portion of the representative element shown by dotted box above). Here $dg$ signifies the groove depth in relation to the reference state (red dotted line) and $\Upsilon$ denotes the angle formed between the tangent and the horizontal axis at the groove's root.}
}
\label{fig:evol}
\end{figure*}

In scenarios where the magnetic field aligns parallel to the Z-axis, we witness a growth of Grain 1 while Grain 2 shrinks as displayed in Figs.~\ref{zoomed}-~\ref{free3}. We anticipate an opposite phenomenon when the field is in the other direction, leading to the shrinkage of Grain 1 and the growth of Grain 2. Upon alignment of the c-axis with the direction of the applied magnetic field, it typically yields lower energy compared to neighboring grains, rendering it energetically suitable. This energy discrepancy prompts the motion of the grain associated with the lower energy contribution, thus displacing the higher-energy grain in the process~\cite{10.1007/978-981-99-6863-3_9}. For more detailed 
information regarding this preferential growth of the grains due to magnetic field, please refer to the works 
by Liang \textit{et al.}~\cite{liang2022phase} and Bandyopadhyay \textit{et al.}
~\cite{10.1007/978-981-99-6863-3_9}. We observe a symmetric groove profile at the initial time step (marked by black in Fig.~\ref{fig:grv3a}). 
However, as the system evolves, the grain boundary migrates due to the magnetic field, and the groove 
associated with the mobile boundary moves toward the right, always displaying an asymmetric profile (marked by 
blue in Fig.~\ref{fig:grv3a}). 

\textcolor{black}{Fig.~\ref{fig:grv3b} zooms into a specific region (marked by a black dotted area) of the groove profile in Fig.~\ref{fig:grv3a}, offering a closer look at its steady-state scenario. Notably, the asymmetry of the groove, results in a significant accumulation of mass in the direction of grain boundary migration. The dashed line in Fig.~\ref{fig:grv3a} represents the mid-plane of the film and the GB position is calculated by tracking the markers (shown by black triangles) along the mid-plane.}

\textcolor{black}{Fig.~\ref{fig:grv3b} illustrates the schematic representation of the grain boundary groove, accompanied by key groove characteristics. In this depiction, $dg$ signifies the groove depth in relation to the reference state, delineated by the red dotted line signifying a flat surface. Additionally, $\Upsilon$ denotes the angle formed between the tangent and the horizontal axis at the groove's root.} 

%Furthermore, the groove formed in the GB-free surface intersection becomes asymmetrical~\cite{verma2022computational} as the groove remains attached to the boundary when it migrates to the right(shown in Fig.~\ref{fig:evol}) under an external magnetic field. 

%Thus, we can attribute that the grain with the c-axis parallel to the direction of the magnetic field grows preferentially at the expense of the other grain with a c-axis perpendicular to the magnetic field direction.refer

In the scenario of a symmetric groove, mass accumulation typically occurs evenly on either side of the boundary and free surface confluence. Upon the motion of the GB with a velocity $\textbf{v}$ from left to right, the x-component of surface-atom ﬂux $J_x$ lodging in the left of the groove root is higher compared to that in the right. Consequently, these atoms migrate significantly further towards the left in comparison to the right, resulting in a notable disparity in mass accumulation adjacent to the moving groove root, hence favoring the accumulation of mass along the direction of the GB migration. Conversely, on the other side of the groove root, corresponding to higher atomic flux, mass accumulation is minimal, thereby inducing an asymmetrical groove profile associated with the migrating grain boundary. For more detailed information regarding the asymmetry of the mobile groove and the concept of flux balance on both sides of the groove root, please refer to the works by Verma \textit{et al}.~\cite{verma2022computational}.

\begin{figure*}[htbp]
  %  \subfloat[groove position at different time steps]{\label{zoomed1}\includegraphics[width=.5\linewidth]{grv_T.jpg}}\hfill
\subfloat[Grain boundary position vs. time]{\label{free21}\includegraphics[width=.49\linewidth]{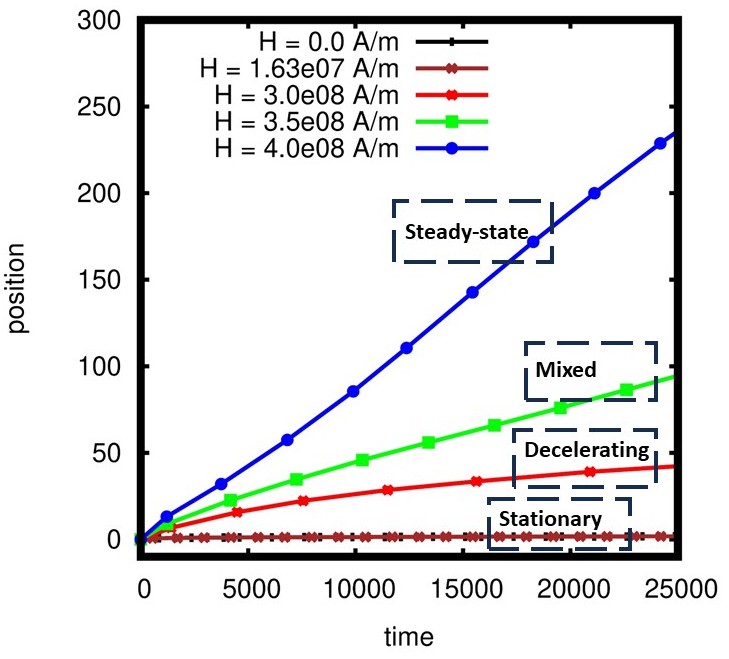}}
\subfloat[Grain boundary velocity vs. time]{\label{free31}\includegraphics[width=.51\linewidth]{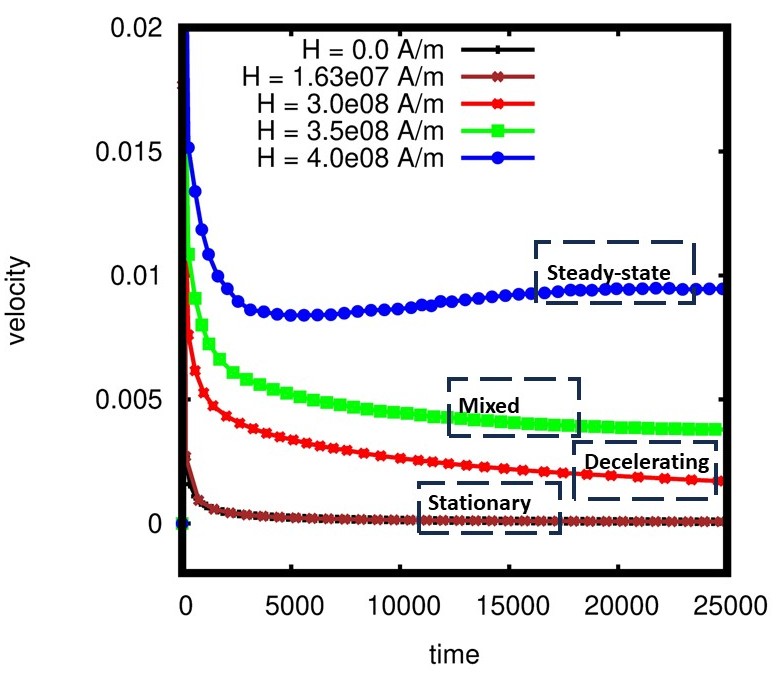}}
%\subfloat[Groove depth vs. time different magnetic fields]{\label{dg_com}\includegraphics[width=.5\linewidth]{dg_all.jpg}}
\caption{Different regimes corresponding to the migration of the mobile groove for different magnetic fields.}
\label{fig:position}
\end{figure*}

Fig.~\ref{free21} shows the grain boundary position vs. time for various applied magnetic loads ($\mathbf{H_\parallel}$ to the Z-axis) in a free-standing thin film bicrystal morphology. We observe a stationary type behavior for lower magnetic field values, i.e., $\textcolor{black}{\mathbf{H} \leqslant 1.63\times 10^8 A/m}$. 
However, a subsequent increase in the magnetic load $\mathbf{H}$ till $3.0\times 10^8 A/m$ displays a decelerating behavior. When we increase the applied magnetic field to $\mathbf{H}=4.0\times 10^8 A/m$, we notice a linear response similar to that observed in the case of steady-state GB motion. The system shows a mixed type of behavior when the applied field is $\mathbf{H}=3.5\times 10^8 A/m$. These trends are also evident in the velocity vs. time profiles for all the cases, as shown in Fig.~\ref{free31}.

%Comparison between (a) grain boundary position vs. time, (b) grain boundary velocity vs. time for different magnitudes of magnetic fields.

\begin{figure*}[htbp]
\subfloat[Without magnetic field]{\label{dg4nomag}\includegraphics[width=.5\linewidth]{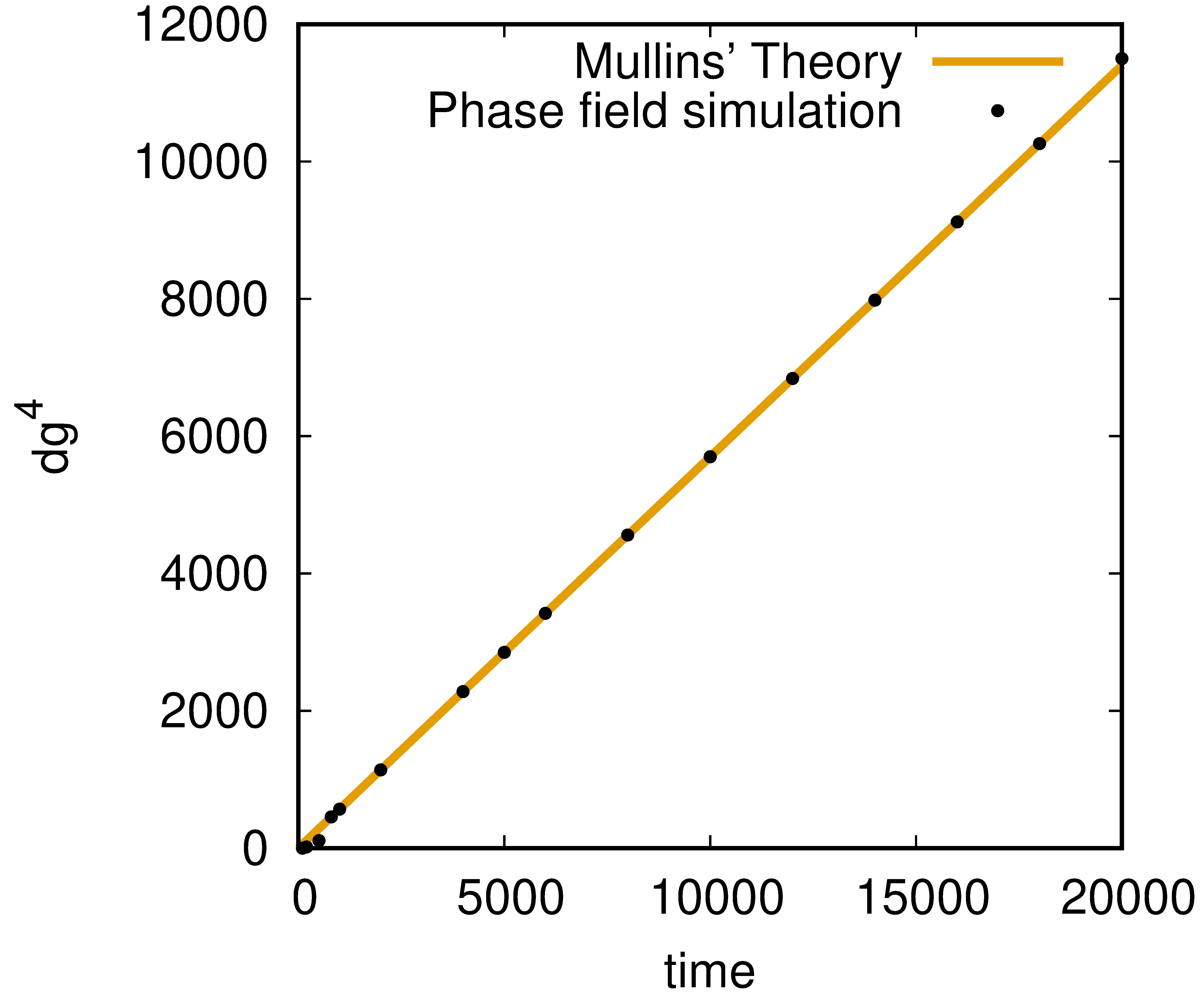}}\hfill
\subfloat[With external magnetic fields]{\label{dg4tmag}\includegraphics[width=.5\linewidth]{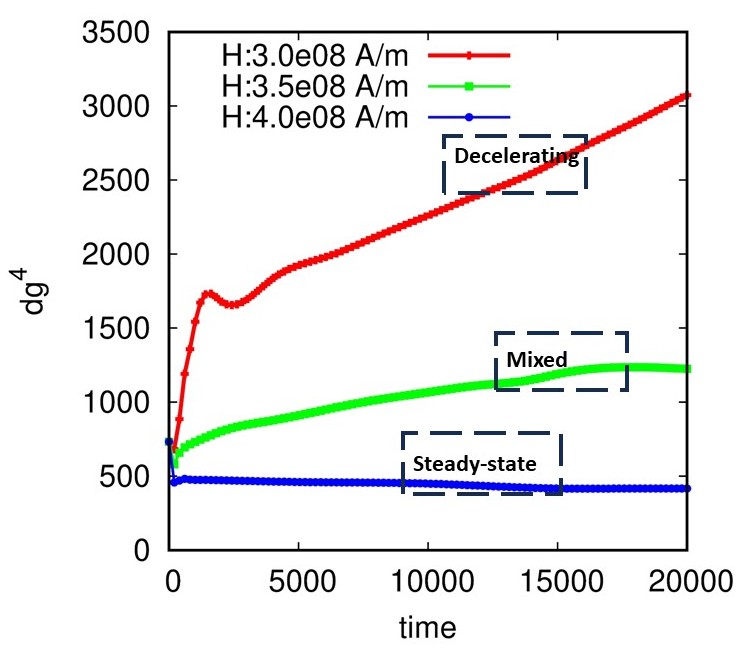}}\hfill 
\caption{Plot depicting the evolution of $dg^4$.}
\label{fig:dg4_t}
\end{figure*}

We investigate different mobile groove behavior subjected to different magnetic fields ($\mathbf{H}=3.0\times 10^8 A/m,3.5\times 10^8 A/m$, and $4.0\times 10^8 A/m$) and compare with the case of zero magnetic field, as illustrated in Fig.~\ref{fig:dg4_t}). In the latter case, the temporal evolution of the fourth power of groove depth $dg^4$ for a stationary boundary follows a linear profile that agrees well with Mullins' analytical solution~\cite{mukherjee2022grain}. Plotting $dg^4$ vs. time corresponding to the 
above-mentioned cases, we observe a continuous increase in the groove depth for a magnetic load of $\mathbf{H}=3.0\times 10^8 A/m$, displaying a decelerating or non-steady state motion. However, for  $\mathbf{H} = 4.0\times 10^8 A/m$, the system displays a steady state profile (marked by blue), and the groove depth remains unchanged at later time steps. On the other hand, when the applied load is $\mathbf{H} = 3.5\times 10^8 A/m$, the system displays a non-steady state behavior for a significant time and tries to display a near steady-state behavior at the later stages. 

%Since there is an analytical solution displaying the characteristics of the steady state behavior of the mobile groove developed by Mullins, we try to validate the steady state behavior obtained by our simulation with Mullin's analytical solution~\cite{mullins1957theory, mullins1958effect}. Information corresponding to the stationary behavior can be obtained in the studies by Mukherjee \emph{et al.}~\cite{mukherjee2022grain}. Although the non-steady state groove behavior and the surface profiles are discussed by Verma \emph{et al.}~\cite{verma2022computational}, they have conducted their studies by assuming a difference in bulk free energy between two grains without any explicit external field. In this study, we also obtain similar decelerating behavior (time-independent groove behavior) and self-similar surface profiles in the presence of an external magnetic field (See Supplementary section~\ref{sec:sup1} ). 

\begin{figure*}[!htbp]
\subfloat[]{\label{zoomed2}\includegraphics[width=0.33\linewidth]{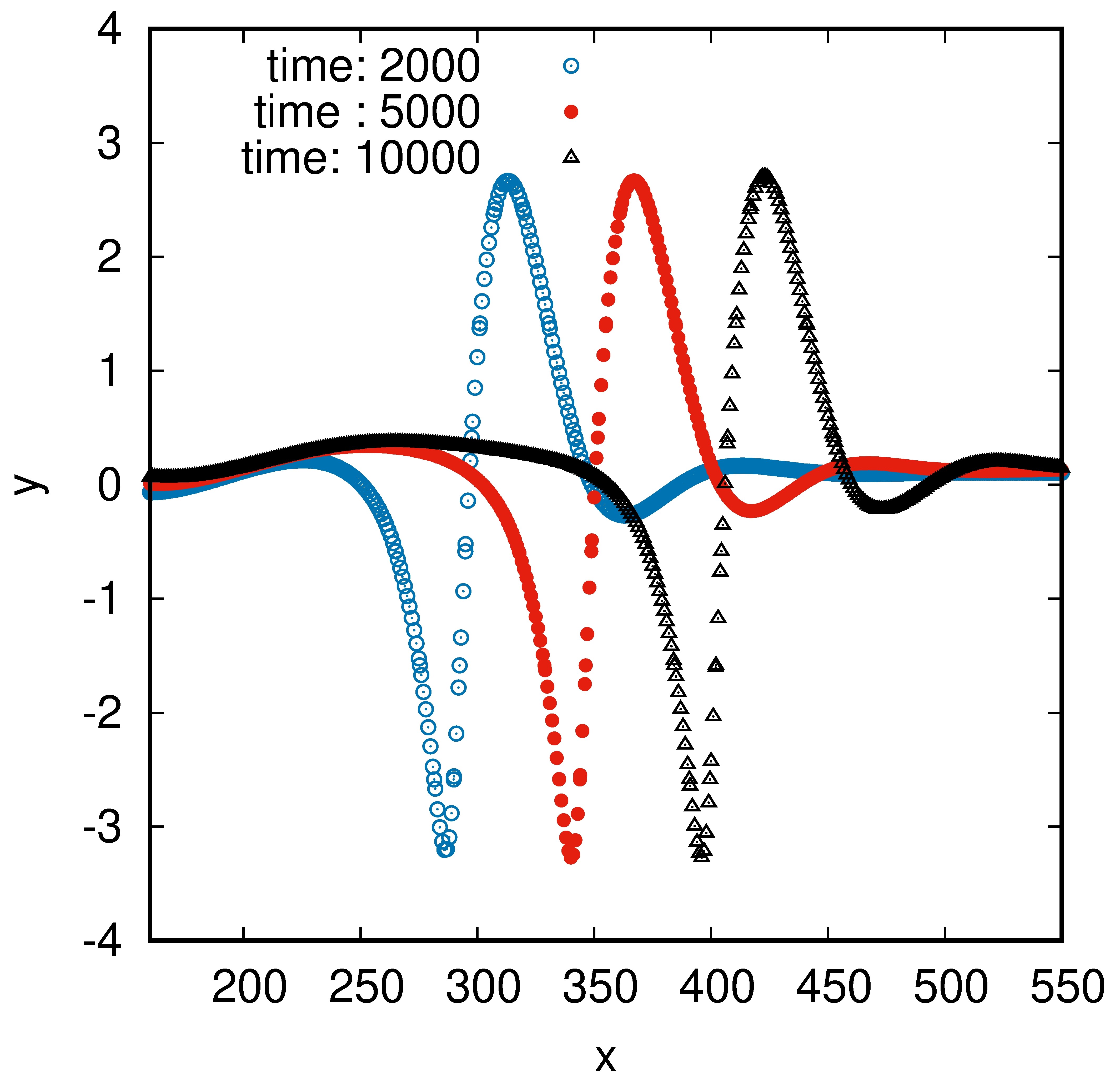}}
\subfloat[]{\label{free22}\includegraphics[width=0.33\linewidth]{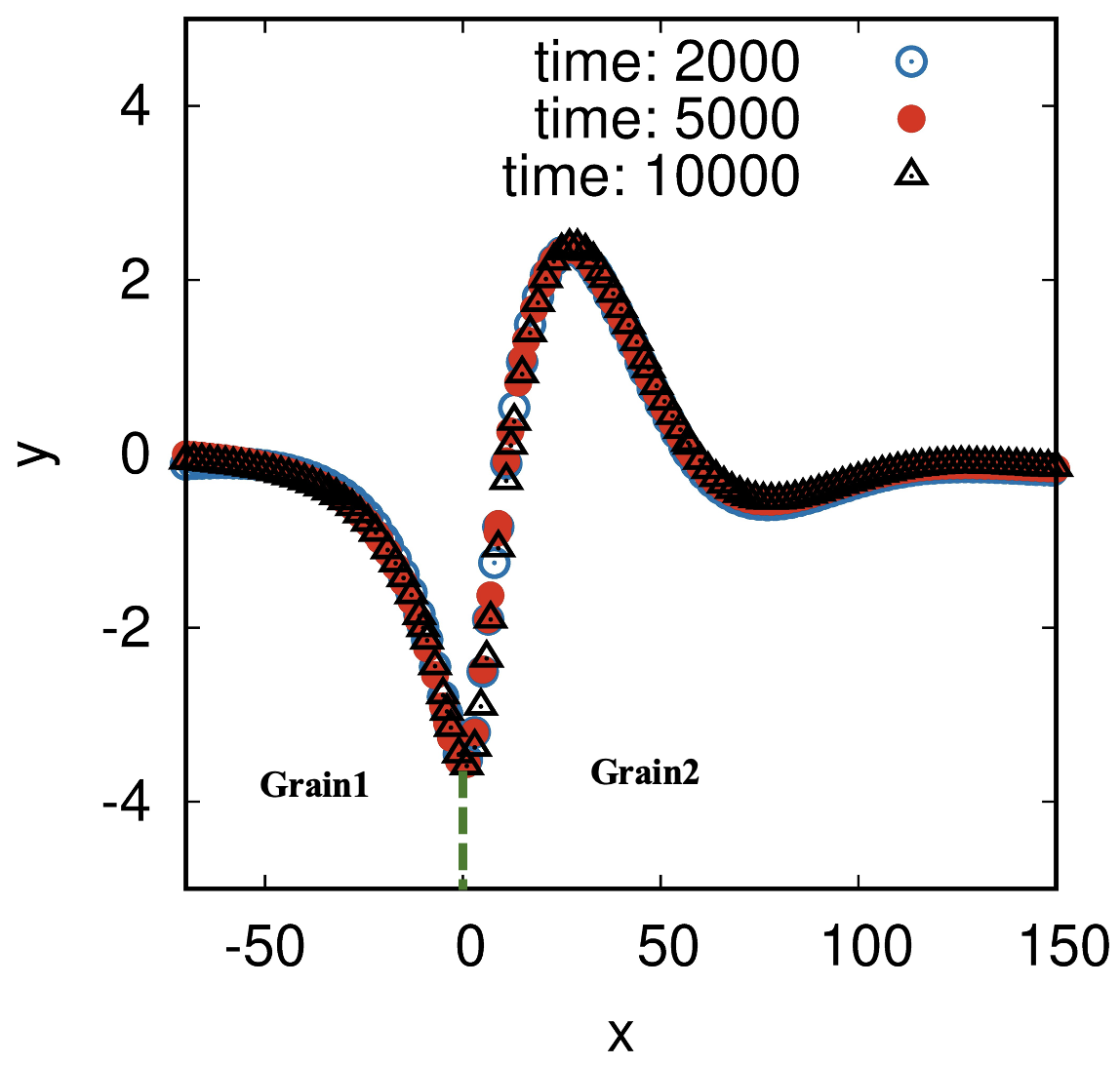}}
\subfloat[]{\label{free32}\includegraphics[width=0.35\linewidth]{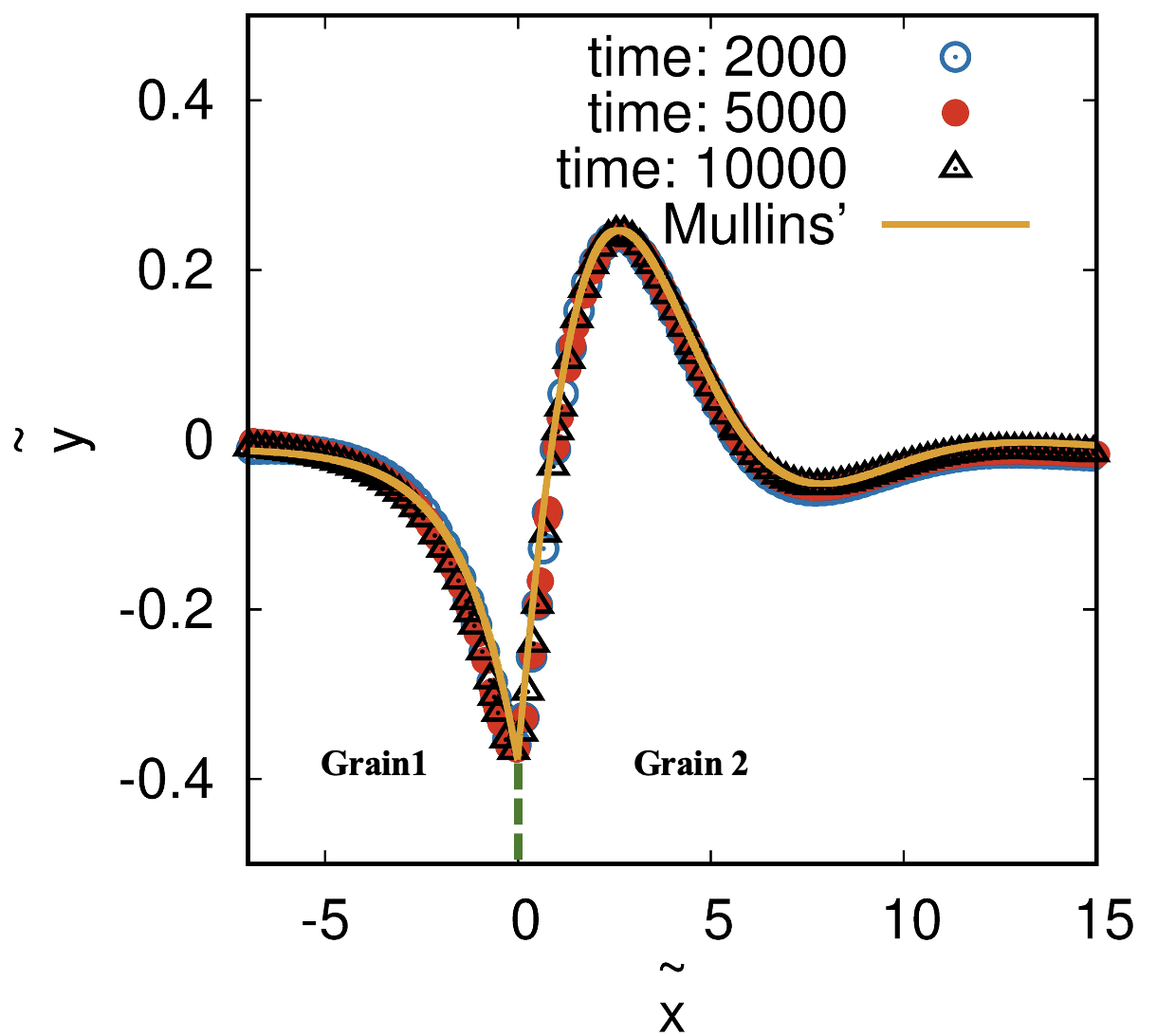}}
\caption{(a) Surface profiles of a mobile groove for $H_m = 4.0\times 10^8 A/m$. (b) Surface profiles with shifted origin. (c) Normalized surface profiles. Surface profiles with shifted origins show complete overlap. The normalized profiles exhibit self-similarity and closely align with the analytical description of Mullins. \textcolor{black}{Here, the dotted line signifies the grain boundary between the grains.}}
\label{fig:surface profiles}
\end{figure*}

Fig.~\ref{zoomed2} illustrates the progression of the surface proﬁle of a migrating groove over time in a free-standing film subjected to applied magnetic field $\mathbf{H} = 4.0\times 10^8 A/m$. Since the system displays a steady state response at this magnetic field, we also try to validate our results (steady state surface profiles) with that analytically obtained by Mullins. We also observe the profiles to be self-similar when we shift the origin of the surface profiles(Fig.~\ref{free22}) to the groove root. In order to compare with Mullins' analytical solution, we normalize these proﬁles relative to the groove depth. Remarkably, the normalized profiles also align with the analytical description provided by Mullins~\cite{mullins1957theory}, evident in Fig.~\ref{free32}. Note that we use a factor \textcolor{black}{$\delta = dg/(0.78tan\Upsilon)$}, where $\Upsilon$ represents an angle formed by the groove with the horizontal axis and $dg$ corresponds to the groove depth, to scale both the horizontal and vertical axes in order to plot the normalized surface profiles $\tilde{y}(\tilde{x})$. 

The profiles obtained from our simulations exhibit a remarkable quantitative alignment with Mullins' analytical solution. In this case, for an applied magnetic field of $\mathbf{H} = 4.0\times 10^8 A/m$, the value of $\textcolor{black}{\gamma_{GB}}/\gamma_s$ is $0.76$. 
Although no proper analytical solution exists for the non-steady state behavior of the mobile grooves, they also exhibit self-similar surface profiles (See Supplementary material~\ref{sec:sup1} for more information related to the non-steady state mobile groove.)~\cite{verma2021grain}. 

\subsection{Three-dimensional simulations of grain boundary grooving under applied magnetic field}
\label{3d}
The above section mainly focuses on the different behavior of a mobile groove under magnetic load and to study such groove profiles and their behavior at the GB, 2D simulations are adequate to explain the underlying physics; however, in a real polycrystalline scenario, where multiple junctions (triple point, quadruple points, etc.) coexist along with the GB's the interaction becomes complex due to combined magneto-chemical synergism. Moreover, in addition to the formation of grooves at the GBs, a polycrystalline thin film with a free surface tends to form pits at the triple junctions, which interact with one another~\cite{saidov2023direct}. To investigate and understand such interactions, we need to perform three-dimensional simulations. Thus, in this section, we extend our model into 3D to study the complex groove-pit interplay under an applied magnetic field.

\begin{figure*}[!htpb]
\includegraphics[scale = 0.25]{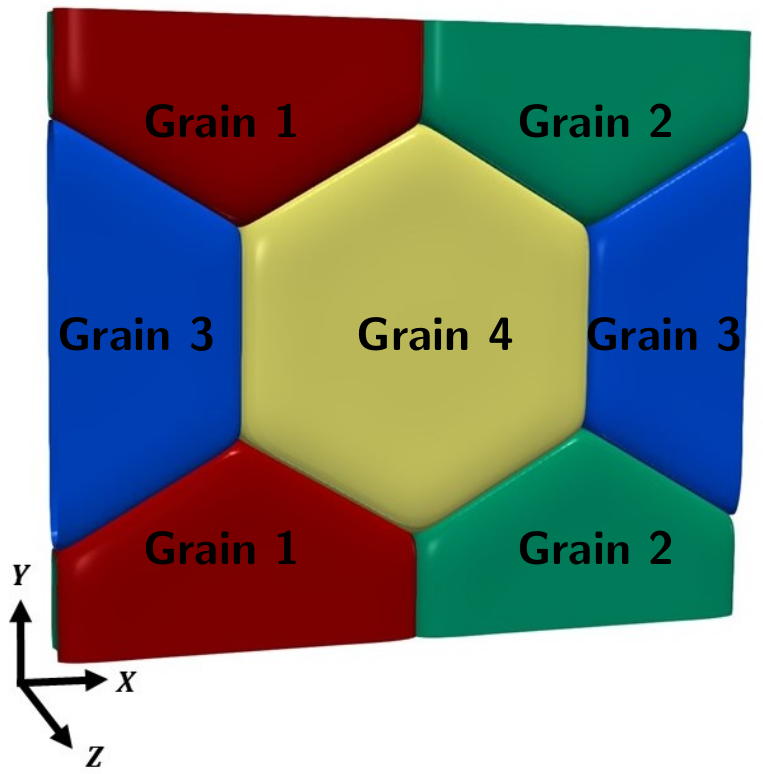}
\caption{ Initial 3D microstructure of an equal-sized hexagonal polycrystal. The system consists of four grains, which are marked with different IDs.}
\label{fig:3dfin}
\end{figure*}

We consider a 3D system of dimensions $200\Delta x \times 174\Delta y \times 30\Delta z$ representing an ideal polycrystalline scenario comprising nearly equisized columnar hexagonal grains. In the 
direction of the Z-axis, the film possesses a material 
thickness of $20\Delta z$, with a vapor phase extending $5\Delta z$ 
on each side at the top and bottom. Additionally, periodic boundary conditions are employed along all three dimensions. It is important to note that the size of the vapor phase is sufficiently large to disregard interactions between the solid thin films across the boundary, given the imposed periodicity. Fig.~\ref{fig:3dfin} represents the microstructure of the studied system where we have taken four grains and marked different grains with different IDs (Grain 1, Grain 2, etc.). 

\begin{figure*}[!htpb]
\subfloat[]{\label{Pitting3d}\includegraphics[width=0.4\linewidth]{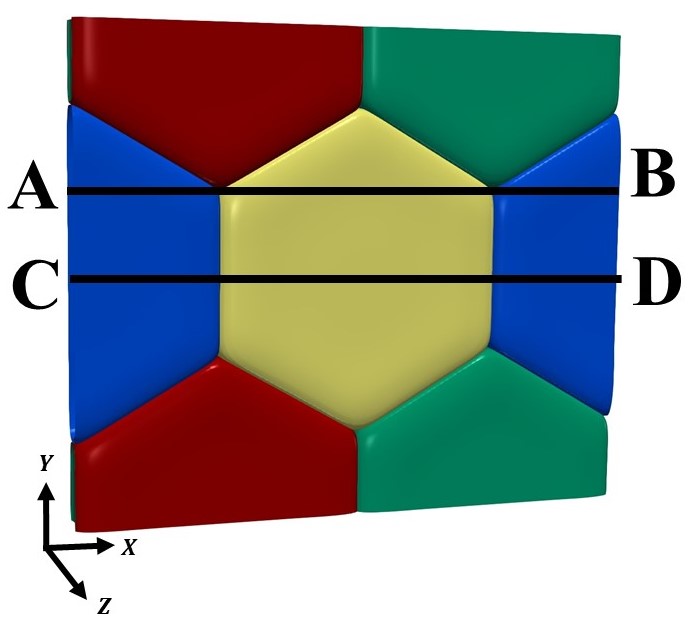}}\hfill
\subfloat[]{\label{surfaceprof3d}\includegraphics[width=.55\linewidth]{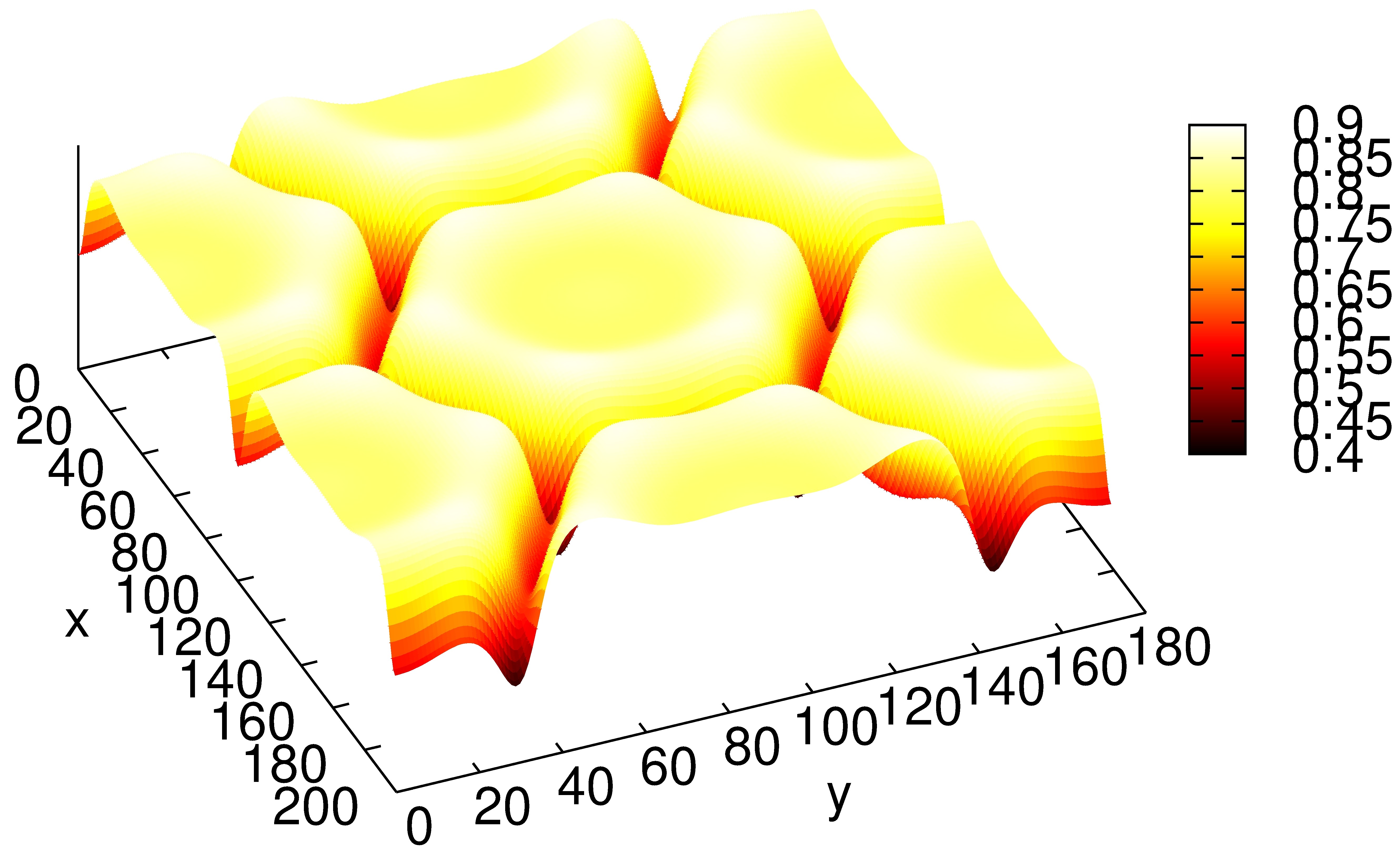}}\hfill 
\subfloat[]{\label{surfaceprof}\includegraphics[width=.4\linewidth]{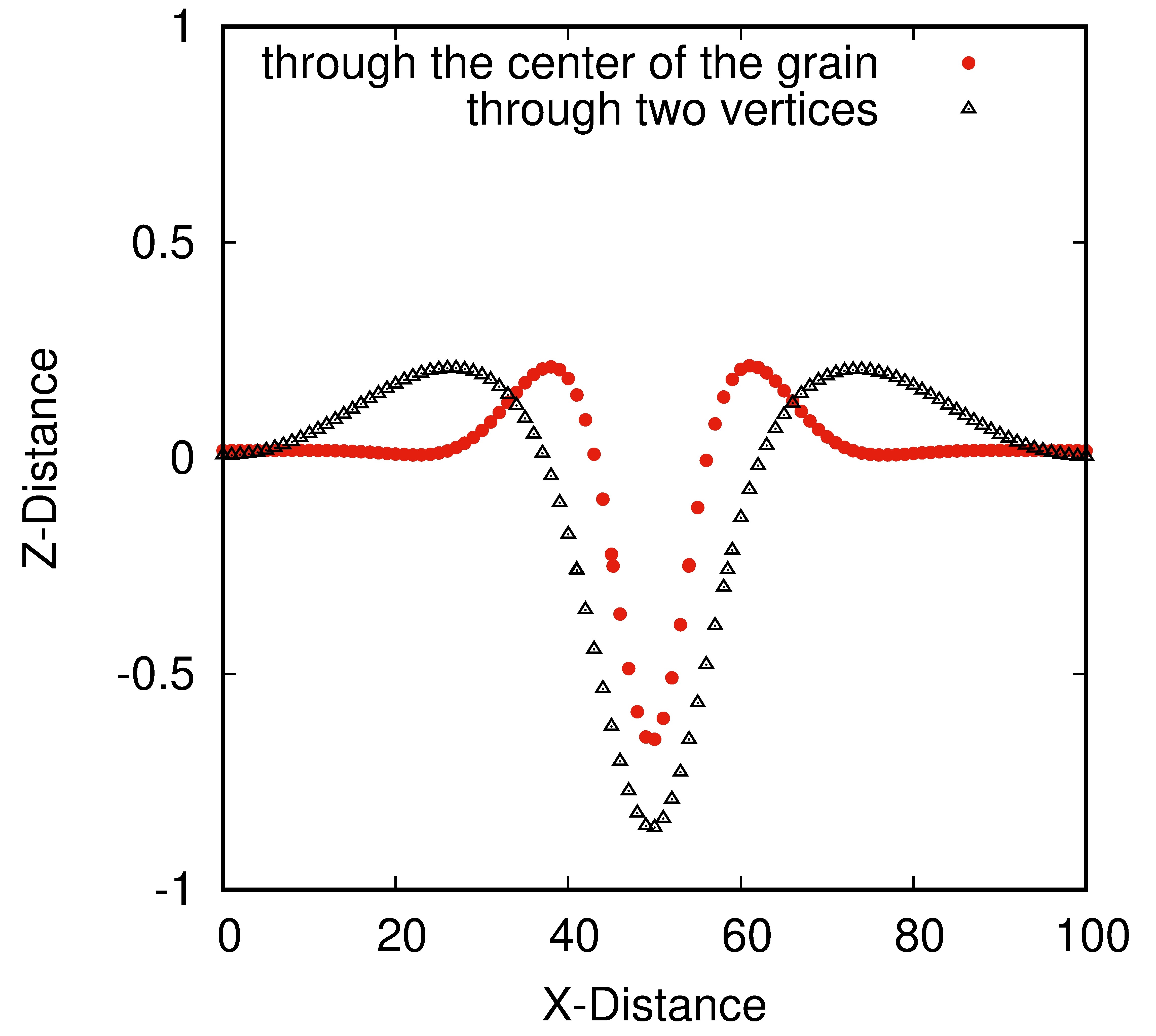}}\hfill 
\subfloat[]{\label{ratio_wo_mag}\includegraphics[width=.43\linewidth]{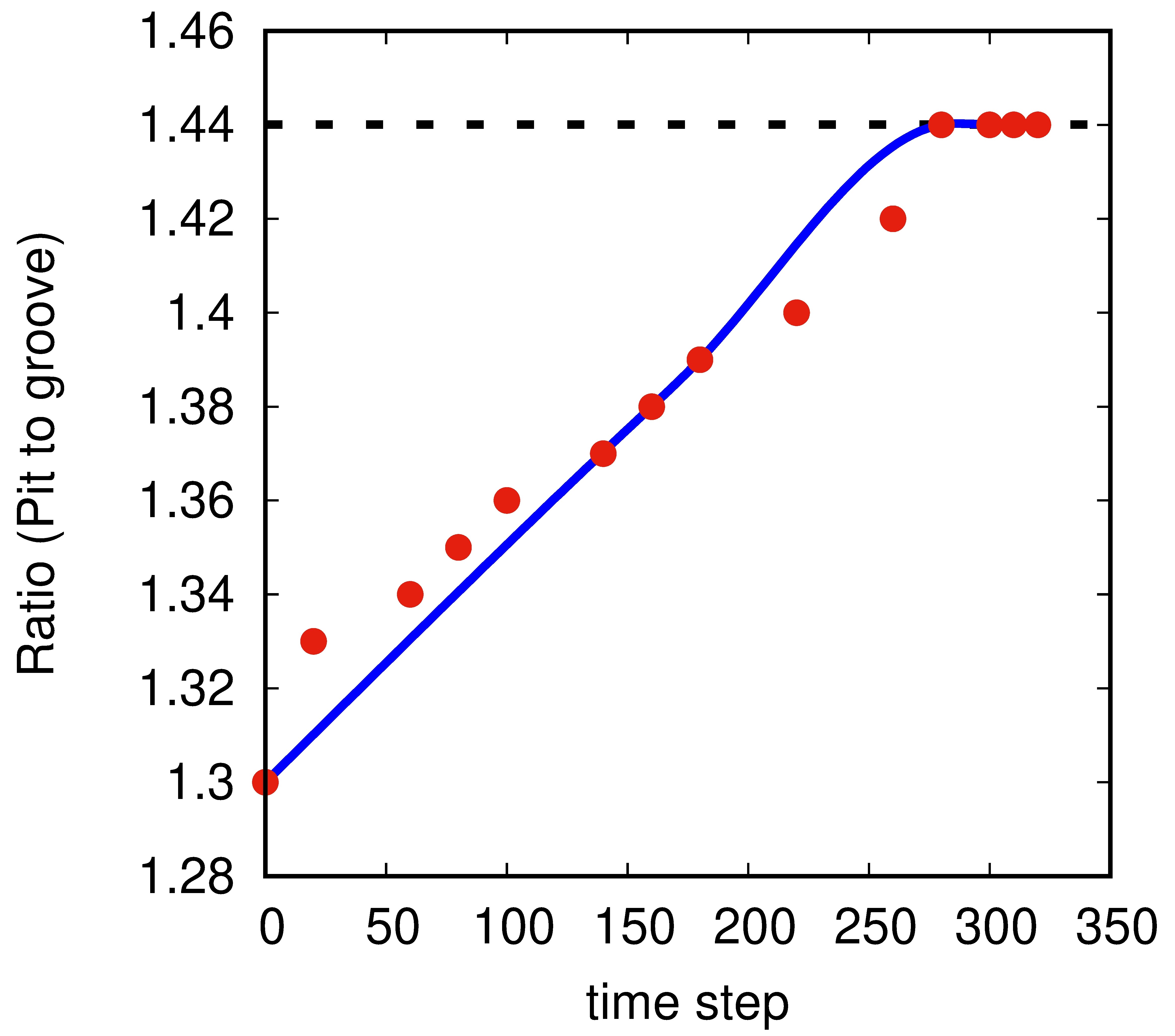}}
\caption{(a) Microstructure with four identical sized columnar hexagonal grains without magnetic field in 3D. 
\textcolor{black}{(b) Three dimensional representation of the surface profile showing distinct grooves and pits for the columnar hexagonal grains}.  
(c) Surface profiles taken perpendicular to the opposing grain boundaries, specifically along AB and CD, respectively\textcolor{red}{.} (d) The $dp/dg$ is plotted with respect to time.  \textcolor{black}{The fitted blue line serves as a guide to the eyes.}}
\label{fig:3Dprofiles}
\end{figure*}
 
Fig.~\ref{Pitting3d} displays a 3D microstructure of the final state of the representative system (for $\mathbf{H} = 0 A/m$) where the grain groove is clearly visible. \textcolor{black}{We also plot the three dimensional representation of the surface profile for the mentioned system in Fig.~\ref{surfaceprof3d}, where the formation of distinct grooves at the boundaries and pits at the triple junctions along with the hills and valley regions on the surface are clearly visible.}
Moreover, we depict surface profiles (illustrated in Fig.~\ref{surfaceprof}) along lines perpendicular to two opposing grain boundaries. One line passes through the grain-center, while the other is traced through two vertices (designated as lines \textcolor{black}{CD and AB} in Fig.~\ref{Pitting3d}). 
The comparison between grooving at the vertices and that at the boundary, as illustrated in Fig.~\ref{surfaceprof}, clearly indicates a more pronounced depth at the vertices.
This typical phenomenon is known as pitting~\cite{mukherjee2022grain}. 
Fig.~\ref{ratio_wo_mag} illustrates the temporal evolution of the pit depth ($d_p$) to groove depth ($d_g$). Notably, there is an initial rise in the ratio followed by stabilization around 1.44. This suggests that in 3D systems, the phenomenon (grooving and pitting) can be characterized by a unified length scale~\cite{mukherjee2022grain}. This also implies that $dp^4$ and $dg^4$ display a linear response as a function of time; however, the former should exhibit a greater slope than the latter one~\cite{mukherjee2022grain}. In a recent qualitative study of grooving in a three-dimensional hexagonal system, Mukherjee \textit{et al.} reports a similar phenomenon~\cite{mukherjee2022grain}. In the following section, we apply a magnetic load in this ideal polycrystalline hexagonal system and try to explore the combined groove-pit interaction in the presence of an applied external magnetic field.  

%\subsubsection{\bf{Effect of applied magnetic field}}
%\label{magnetic}
%We have shown in Section~\ref{sec:formula} the form of the magnetic free energy density for a three-dimensional scenario (Eq.~\eqref{magE3d}), where we already discussed the angles $\theta$ and $\phi$ between the applied magnetic field $\mathbf{H}$ and ND as well as RD.
%We assume two different grain orientations, $0^\circ$ for Grain 1, Grain 2, Grain 3 and $90^\circ$ for Grain 4, such that we obtain maximum anisotropy in the magnetic driving force similar to the condition considered in the case of the bicrystal system discussed previously. 

\begin{figure*}[!htpb]
\includegraphics[scale = 0.5]{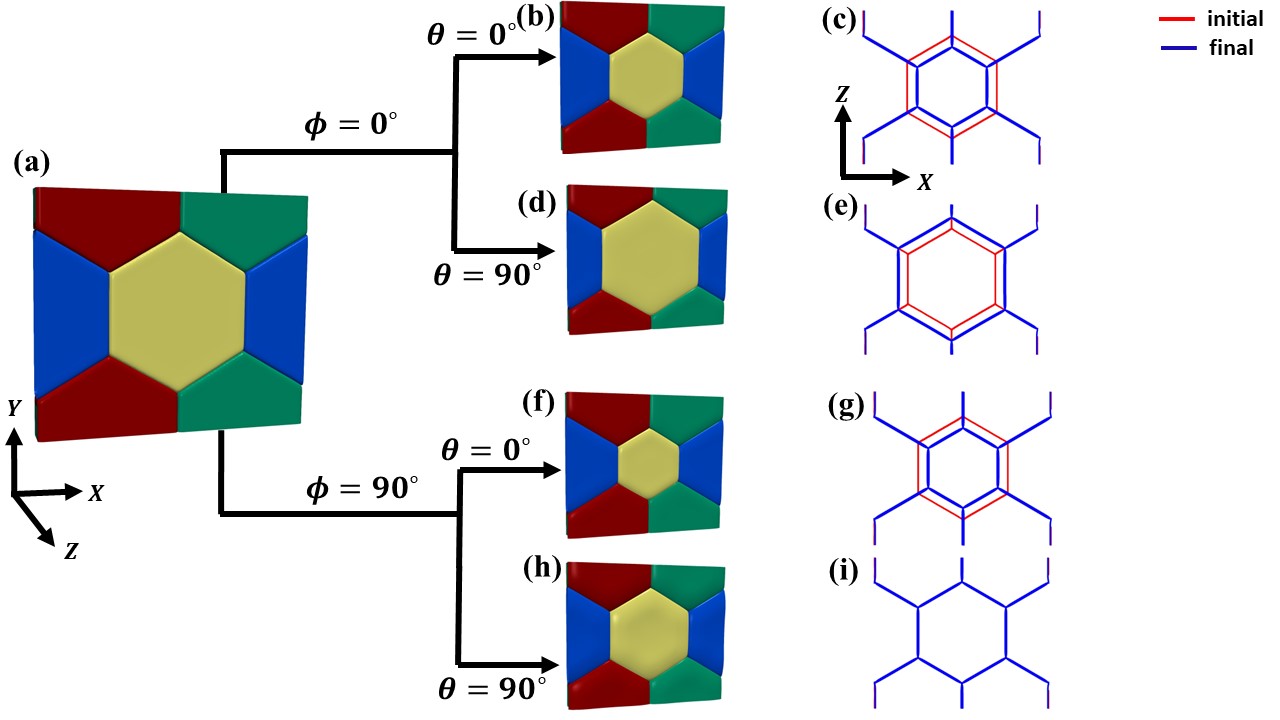}
\caption{(a) Initial microstructure with equal-sized grains. (b), (f) Final microstructures and (c), (g) contour plots for case 1a and case 2a, showing the shrinkage of Grain 4 at the expense of others. (d) Final microstructure and (e) contour plot for case 1b, showing the growth of Grain 4 at the expense of others. (h) Final microstructure and (i) contour plot for case 2b, where the morphology remains unchanged, similar to zero magnetic field.}
\label{fig:hexamag3d}
\end{figure*}

We assume Grain 1, Grain 2, and Grain 3 possess the c-axis aligned parallel to the Z-direction. In contrast, the crystallographic $c$-axis of Grain 4 is oriented perpendicular to the Z-axis. It can be parallel to the $X$ or the $Y$ axis, and we consider the former direction. As previously noted, this orientation guarantees the highest degree of anisotropy in the magnetic energy. We investigate two different cases:
\begin{enumerate}
\item case 1: $\phi = 0^\circ$
      \begin{itemize}
          \item case 1a: $\theta = 0^\circ$
          \item case 1b: $\theta = 90^\circ$  
      \end{itemize}
      \item case 2: $\phi = 90^\circ$
    \begin{itemize}
         \item case 2a: $\theta = 0^\circ$ 
         \item case 2b: $\theta = 90^\circ$ 
\end{itemize}           
\end{enumerate}
We use the same notation for $\theta$ and $\phi$, as discussed in Section~\ref{sec:formula}.

Fig.~\ref{fig:hexamag3d}(a) shows the initial configuration of the system considered for all the cases.  
As the system evolves (case 1a), we observe Grain 4 shrinks and the other grains increase, maintaining the ideal hexagonal morphology as displayed in Fig.~\ref{fig:hexamag3d}(b). This phenomenon is similar to that observed in the case of the bicrystal study, and we can provide a similar explanation in terms of energy.
\textcolor{black}{Here, depending on the orientation of the magnetic field and the alignment of the $\textrm{c-}$axis of the corresponding grain the total energy of the center grain (Grain 4) can be higher or lower compared to it's neighboring grains.}
In this case, Grain 4 possesses higher energy than others, which sets off the preferential growth of Grains 1-3. The corresponding contour plots of the initial and final microstructure (shown in Fig.~\ref{fig:hexamag3d}(c)) display the shrinkage of Grain 4 at the expense of others (See Supplementary section~\ref{sec:sup2}). We observe an exact opposite phenomenon for case 1b. In this case, Grain 4 grows at the expense of the other grains (Fig.~\ref{fig:hexamag3d}(d)). Fig.~\ref{fig:hexamag3d}(e) shows a contour plot displaying the initial and final configurations. For case 2a (shown in Figs.~\ref{fig:hexamag3d}(f)-~\ref{fig:hexamag3d}(g)), we observe a behavior similar to case 1a, where Grain 4 shrinks and others grow.

However, an intriguing scenario occurs for case 2b, i.e., $\phi:90^\circ$ \& $\theta:90^\circ$ (Fig.~\ref{fig:hexamag3d}(h)). Fig.~\ref{fig:hexamag3d}(i) displays the contour plots of the initial and final configurations for case 2b. In this particular case, we do not observe any growth behavior of the grains, and it reflects a scenario similar to the case without an applied magnetic field. The extra energy in the grains created due to the magnetic field neutralizes the anisotropic contribution in the energy resulting from the difference in the susceptibility in the $Z$, $X$, and $Y$ directions due to different grain orientations. Table~\ref{Tab:tab1} tabulates the conditions and scenario for both the cases. When the magnetic energy ratio between Grain 4 and other grains is less than 1, Grain 4 shrinks. On the contrary, when it is greater than 1, it grows. However, when it equals 1, all the grains remain unchanged.

\begin{table*}[htb]
    \caption{\textcolor{black}{Phenomena observed under various conditions, with respect to $\theta$ and $\phi$.}}
    \small
\begin{tabularx}{\linewidth}{|p{3.4em}
                             |l
                             |l
                             |X
                             |X|}
    \Xhline{0.8pt}
\thead{$\phi$}
    & \thead{$\theta$}  & \thead{Energy ratio (Approx.)\\(\textcolor{black}{Grain 4 /Grains 1-3}})    & \thead{\textcolor{black}{Grains 1-3}}  & \thead{\textcolor{black}{Grains 4}}              \\
    \Xhline{0.8pt}
\multirow{2}{=}{$0^\circ$}
    &   \textcolor{black}{case 1a}: $\theta = 0^\circ$
                &   $ \simeq 0.84$      & grow  & shrink                   \\
    \cline{2-5}
    &   \textcolor{black}{case 1b}: $\theta = 90^\circ$
                &  $ \simeq 1.18$ 
                             
      &shrink 
       & grow        \\
 
    \hline
\multirow{2}{=}{$90^\circ$}
    &  \textcolor{black}{case 2a}: $\theta = 0^\circ$
                &   $ \simeq 0.84$         & grow  & shrink  \\
    \cline{2-5}
    &   \textcolor{black}{case 2b}: $\theta = 90^\circ$
                &  $\simeq 1$                       
                & unchanged  & unchanged   \\
    \Xhline{0.3pt}
\end{tabularx}
\label{Tab:tab1}
    \end{table*}

%\begin{figure}[htbp]
%    \centering
%    \subfloat[Initial]{\label{initialphi90theta90}\includegraphics[width=0.4\linewidth]{initial_phi0theta_0.jpeg}}\hfill
%\subfloat[Final]{\label{finalphi90theta90}\includegraphics[width=.38\linewidth]{final_phi90theta90.jpeg}}\hfill 
%\subfloat[Surface profile after final step]{\label{surf_phi0theta90}\includegraphics[width=.5\linewidth]%{groove_pit_other1_phi90theta90.jpg}}
%    \caption{case 2b: 3D microstructure of thin film with four equal-sized hexagonal grains with applied magnetic field at (a) Initial (b) Final time step, (c) Surface profiles perpendicular to opposite grain boundaries.}
%    \label{fig:phi90theta90mag}
%\end{figure}

\begin{figure*}[!htbp]
%\centering
\subfloat[]{\label{grv_pit_all}\includegraphics[width=0.33\linewidth]{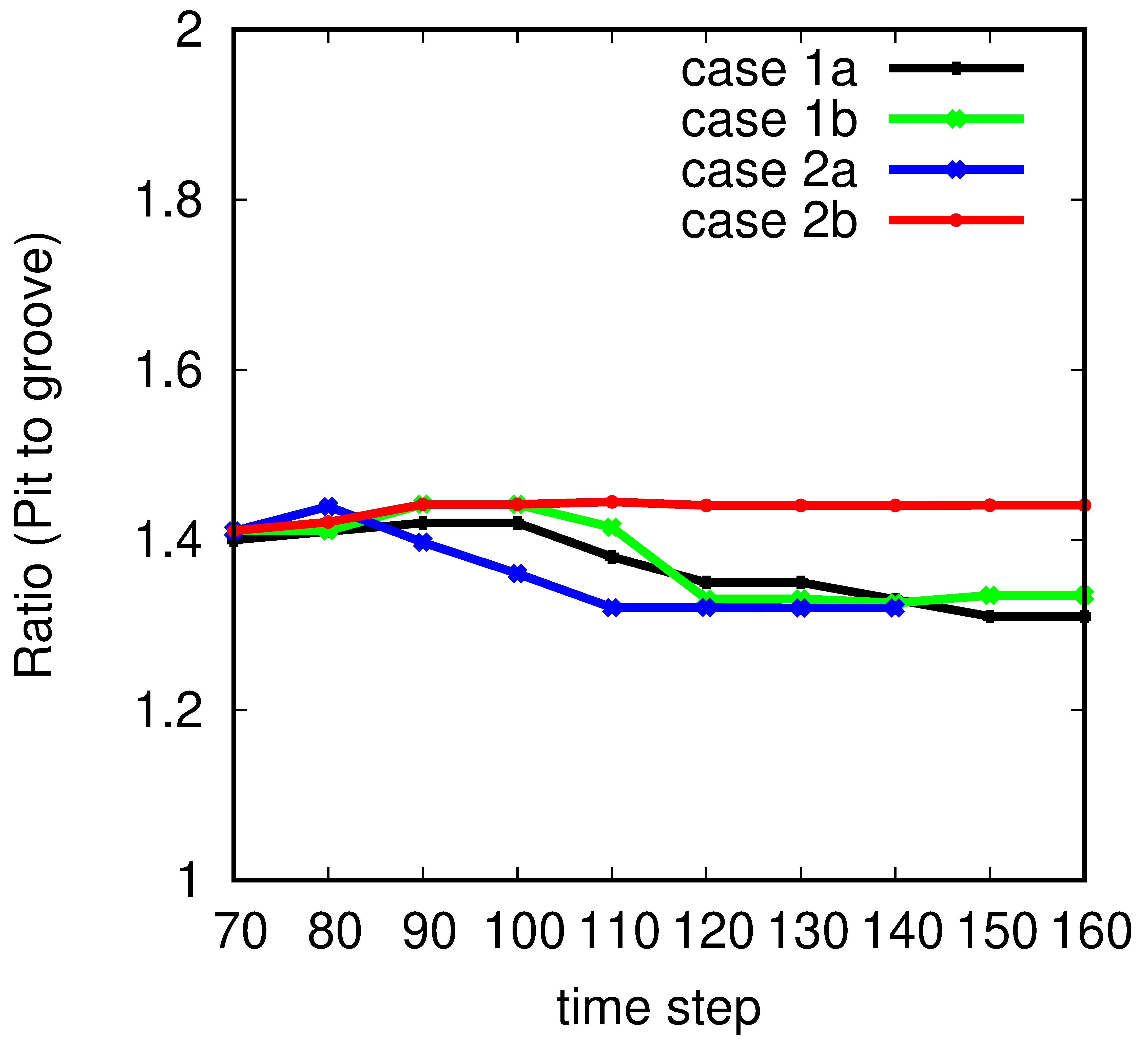}}
\subfloat[]{\label{analy_compare}\includegraphics[width=0.33\linewidth]{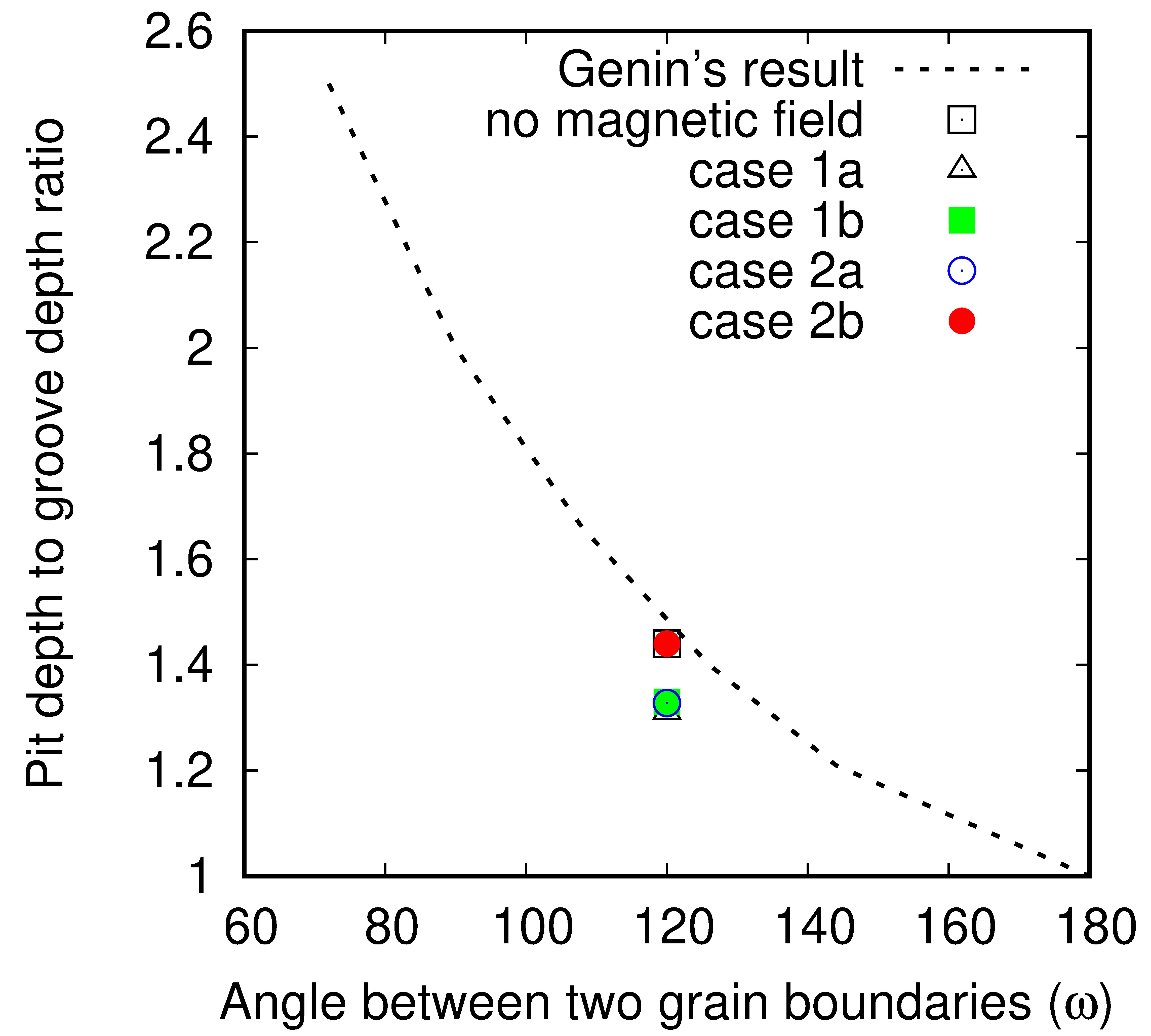}}
\subfloat[]{\label{dp_dg_com}\includegraphics[width=0.32\linewidth]{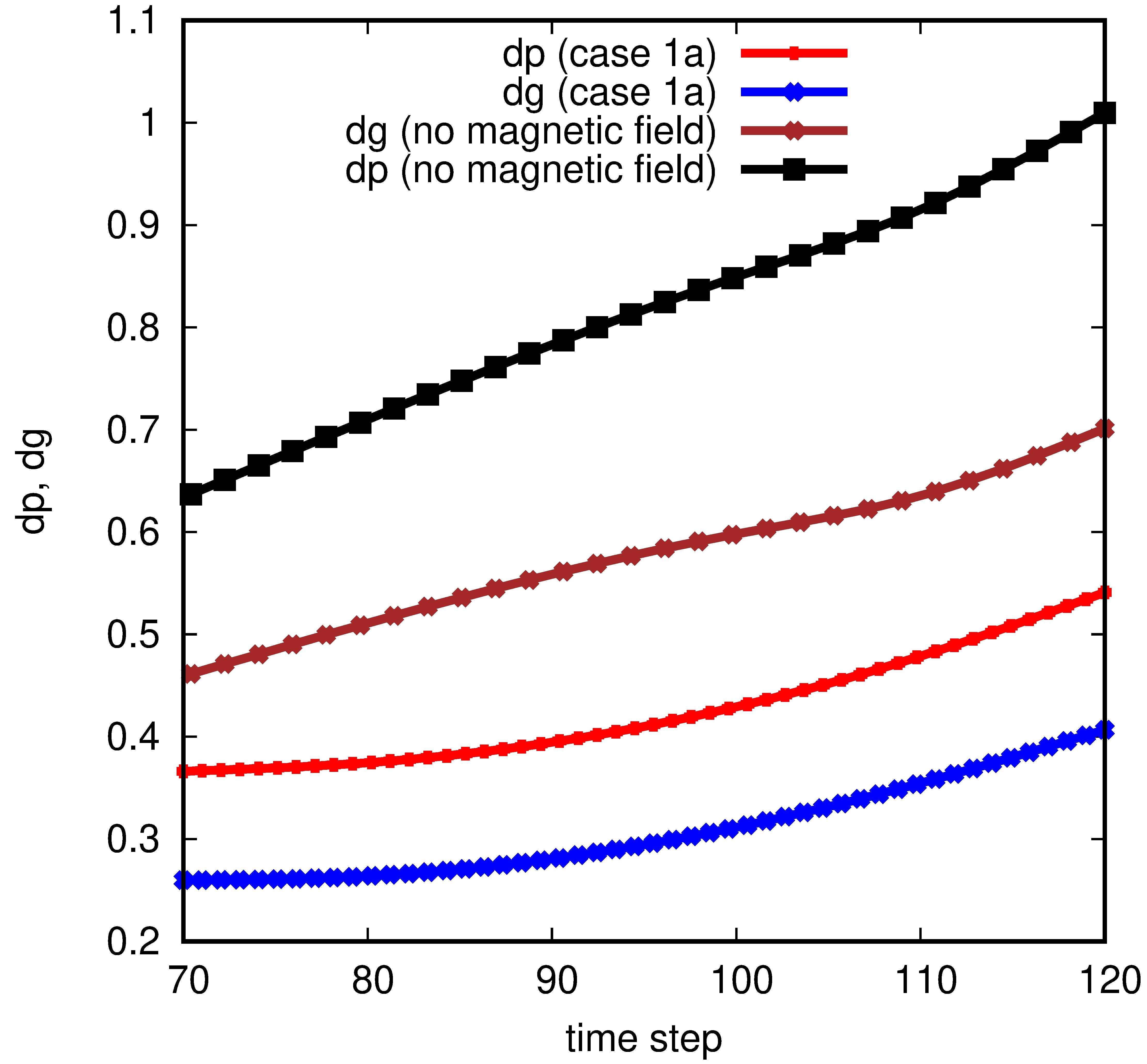}}
\caption{(a) Temporal evolution of $dp/dg$ for different cases of $\phi$ and $\theta$. (b) Comparison of $dp/dg$: phase-field results and Genin \textit{et al.}~\cite{genin1992capillary}. (c) Comparison between the slope of $dp$ and $dg$ for cases without magnetic field and with magnetic field (case 1a). The higher slope reduction in $dp$ compared to $dg$ for case 1a indicates the higher drag experienced at the triple point by the system under external magnetic load.}
\label{fig:grv_pit_compare}
\end{figure*}

Fig.~\ref{grv_pit_all}  shows the ratio of $dp$ and $dg$ over time. For case 2b, we obtain $dp/dg \simeq 1.44$, a value akin to the scenario where no external magnetic field is applied. On the other hand, we observe a decrease in the pit-to-groove depth ratio for all the other scenarios. Moreover, we also compare the depth ratio ($dp/dg$) obtained from our computational analysis of grooving in this case (where the GBs form a $120^\circ$ angle) with the findings of Genin et al.~\cite{genin1992capillary} (shown in Fig.~\ref{analy_compare}) who investigated the emergence of pits at the convergence of two boundaries intersecting at different angles denoted by $\omega$. For the case without magnetic field as well as case 2b, the ratio ($dp/dg$) nearly falls on the analytical profile obtained by Genin for $\omega = 120^\circ$. However, we obtain a lower value $\simeq 1.32$ for all the other cases. Hence, we consider case 1a as our representative system of interest.

%Since, except case 2b, all the other cases display a lower and almost similar pit-to-groove depth ratio irrespective of the direction of the applied magnetic field; we consider case 1a as our representative system of interest.

The drop in the $dp/dg$ ratio can be attributed to the hindered motion of the pits compared to the groove, which is also evident in Fig.~\ref{dp_dg_com}. Here we compare the deepening of the pit ($dp$) and groove ($dg$) between the case with no applied magnetic field and case 1a. Similar behavior is reported in the literature for polycrystalline bulk systems, where the triple junction drag occurs during grain growth~\cite{gottstein2010thermodynamics,gottstein2002triple}. We also observe a prominent decrease in the slope of the pit (slope of $dp$ has reduced to $\approx 47\%$) as well as groove depth (slope of $dg$ has reduced to $\approx30\%$) for case 1a compared to the case where we do not apply a magnetic field. The larger decrease of slope in $dp$ is due to the significant drag force experienced by the pit forming at triple junctions compared to the grooves that form in GBs~\cite{gottstein2002triple}. 

%as a function of the angle ($\omega$) and different cases of $\phi$ and $\theta$ and our  are also compared with the results of
%Since, except case 2b (where we observe a stationary type behavior), all the other cases display a lower and almost similar pit-to-groove depth ratio irrespective of the direction of the applied magnetic field; we consider case 1a as our representative system of interest.
\newpage
\section{Conclusions}
In this work, we have developed a phase-field model to study microstructural evolution and GB migration in thin-film configurations consisting of 
solid film and a vapor phase under a magnetic load applied externally. 
\textcolor{black}{Apart from the curvature driven grain growth, grain boundary (GB) migration is primarily driven by variation in total energy across the grains. Previous researches often relied on hypothetical assumptions regarding the free energy disparity between grains to initiate GB migration qualitatively. In our study, we introduce a quantitative approach by utilizing an external magnetic field to induce GB migration, a method feasible in practical applications. 
Our findings demonstrate that applying an external magnetic field offers tangible control over GB migration and influences groove profiles.} Furthermore, this model allows us to study the complex interaction between surface diffusion-controlled GB grooving and GB 
migration subjected to an external magnetic field in a diamagnetic anisotropic system Bismuth (Bi) with a 
hexagonal crystal structure. \textcolor{black}{This study also discusses the formation and the combined interaction between the grooves (at the GBs) and pits (at the triple junctions) subjected to external magnetic field in polycrystalline films. To our knowledge, no proper study discusses the combined groove-pit interactions due to GB migration under an external magnetic field.}

Thus, the following conclusions can be drawn based on our study.
\begin{itemize}
\item Grains aligned with their crystallographic c-axis parallel to the magnetic field exhibit preferential growth, as they contribute less to the total energy, leading to their dominance over other grains. Differences in free energy between differently oriented grains result from anisotropic susceptibility due to the inherent crystallographic anisotropy of the hexagonal crystals.
\item Our study reveals different regimes of grain boundary motion at different magnetic fields. In the steady-state regime, the normalized asymmetric surface profiles demonstrate an universal  behavior that aligns closely with Mullins’ theory of mobile grooves, showcasing exceptional quantitative conformity.
\item In our investigation of 3D polycrystalline systems featuring uniform hexagonal grains, we propose a method to regulate and control the growth of grains with distinct orientations through manipulation and tuning of the applied magnetic field direction.
\item We also show the pitting phenomenon and the intricate relationship and interaction between the pits and the grooves formed in a thin polycrystalline film with a free surface. Except for a particular case (case 2b, where the energy ratio of the grains is $\simeq 1$), we observe a reduced pit depth to groove depth ($dp/dg$) ratio compared to that without a magnetic field. This reduction in the ratio indicates the higher drag experienced by the pits at the triple junctions compared to the groove, thus hindering the pit motion.
 
\item The model can further be extended to study the effect of a magnetic field on polycrystalline systems
with a large number of grains with different orientations. Thus, it can provide an alternate route to produce a textured polycrystalline system to obtain better material properties by controlling the magnitude and direction of an external magnetic field. 
Moreover, the contribution of surface energy anisotropy can also be incorporated into the model, which can manifest different complex effects on the GB migration in polycrystalline thin films. 
\end{itemize}
%We plan to perform these studies in our future work.
%In the work, we have systematically studied systems with simple initial configurations. However,

\section{Acknowledgements}
The authors acknowledge financial support from the Science and Engineering Research Board, India (Grant No. CRG/2021/003687) and the Institute Post Doctoral Fellowship, IIT Kanpur. Authors acknowledge the National Supercomputing Mission (NSM) for providing computing resources of ``PARAM Sanganak'' at IIT Kanpur, which is implemented by C-DAC and supported by the Ministry of Electronics and Information Technology (MeitY), India, and the Department of Science and Technology (DST), Government of India. The authors also acknowledge the HPC facility provided by CC, IIT Kanpur.

\section{ Data Availability}
 The raw/processed data required to reproduce these findings cannot be shared at this time as the data also forms part of another ongoing study.
\appendix
%\section{Derivation of the magnetic free energy density in 2D and 3D}
%\label{magderive}

 \section{List of parameters used in the simulations}
 \label{param}
 All the parameters used in this study are in their scaled and non-dimensionalized form.
This section provides a non-dimensionalization procedure of the materials model 
parameters used in the simulation. We choose a characteristic length $L_c$, characteristic energy $E_c$, and characteristic time $\tau_c$
to non-dimensionalize all the parameters. 
The interfacial energy of the system is assumed to be
$\sigma_{GB} = \SI{0.155}{\joule\per\meter^2}$~\cite{ocak2008interfacial}. We choose the interfacial width $l_{GB}$ to be $\SI{1.6}{\nano\meter}$, and in the simulations, we always ensure to have a minimum of five grid points. Thus, we obtain the grid 
spacing $\Delta x$ as $\SI{0.32}{\nano\meter}$.

We also calculate the characteristic energy ($E_c$) considering the Boltzman constant $k_B = \SI{1.38e-23}{\joule\per\kelvin}$ and the temperature $T = \SI{225}{\degreeCelsius}$ as:
\begin{equation}
    E_c = k_B T = \SI{7.2864e-21}{\joule}.
\end{equation}

Additionally, the characteristic length ($L_c$) is calculated to be:
\begin{equation}
    L_c = \sqrt{\frac{E_c}{\sigma_{GB}}} = \SI{2.1682e-10}{\meter}.
\end{equation}

 To calculate the characteristic time $\tau_c$ we use the relaxation parameter $L$
used in the simulation as~\cite{liang2022phase}:
\begin{align}
    L = \frac{4m}{3l_{GB}}.
    \label{kinparam}
\end{align}
Here $m = m_{0}exp(-Q/k_BT)$, $m_0 = \SI{1.1e24}{\meter^4\joule^{-1}\second^{-1}}$ is the pre-exponential factor and $Q = \SI{3.38}{\electronvolt}$ defines the activation energy~\cite{molodov1998true}. Thus using Eq.~\eqref{kinparam} we obtain the mobility parameter $L$ to be $\SI{5.3539}{\meter^3\joule^{-1}\second^{-1}}$. Thus, the characteristic time ($\tau_c$) is calculated as:

\begin{equation}
    \tau_c = \frac{L_c^3}{LE_c} = \SI{2.6129e-10}{\second}.
\end{equation}

Next, we calculate the simulation-specific materials parameters. The free energy constant $B$ in Eq.~\ref{eq2} can be calculated as a function of interfacial energy $\sigma_{GB}$ as well as interfacial width $l_{GB}$ as $B = \frac{6\sigma_{GB}}{l_{GB}} = \SI{5.8125e08}{\joule\meter^{-3}}$. Furthermore, the gradient energy coefficient 
related to the grains $\kappa_{\eta}$ is calculated as $\kappa_{\eta} = \frac{3\sigma_{GB}l_{GB}}{4} = \SI{1.86e-10}{\joule\meter^{-1}}$~\cite{liang2022phase}.

The corresponding non-dimensional values of the material parameters used in the simulation
are tabulated as:
\begin{table*}[htpb]
\begin{tabular}{ ||p{12cm}||p{3cm}||  }
 \hline
 \multicolumn{2}{|c|}{List of parameters (non-dimensional values)} \\
 \hline
Model parameter& Value\\
 \hline
 Bulk free energy parameters (A, B, C, $\Gamma$)   & 1, 0.8131, 1, 1.5\\
 \hline
 Gradient energy coefficient for density ($\kappa_\rho$)   & 2\\
 \hline
 Gradient energy coefficient for order-parameters ($\kappa_\eta$)   & 5.535\\
 \hline
 Bulk Atomic mobility ($M_b$) & $10^{-6}$\\
 \hline
 Surface Atomic mobility ($M_s$) & $1$ \\
  \hline
  Grain boundary relaxation coefficient (L) & 1.0\\
  \hline
 \end{tabular}
\end{table*}

The parameters related to magnetic energy density in Eq.~\ref{eq6} are given as:
$\mu_{0} = \SI{1.257e-06}{\newton\ampere^{-2}}$~\cite{asai2003crystal,liang2022phase}, $\chi_{a} = \chi_{b} = -1.24\times 10^{-4}$~\cite{asai2003crystal,liang2022phase}, 
$\chi_{c} = -1.47\times 10^{-4}$~\cite{molodov2010migration}.

\bibliographystyle{unsrt}
\bibliography{scopus}
\end{document}

\begin{frontmatter}
\title{\textbf{Supplementary Material}\\
%Magnetic field induced grain boundary migration in the presence of thermal grooving in a non-magnetic system: A phase-field study 
Thermal Grooving in Thin Non-magnetic Films: Unraveling the Universal Nature Under External Magnetic field
}
\author[a,b]{Soumya Bandyopadhyay}
\author[b]{Somnath Bhowmick\corref{cor}}
\ead{bsomnath@iitk.ac.in}
\author[b]{Rajdip Mukherjee\corref{cor}}
\ead{rajdipm@iitk.ac.in}
\address[a]{Department of Materials Science and Engineering, University of Florida, Gainesville, Florida-32611, United States}

\address[b]{Department of Materials Science and Engineering, Indian Institute of
Technology, Kanpur, Kanpur-208016, UP, India} 
\end{frontmatter}
\section{\label{sec:sup1} Surface profiles corresponding to the decelerating motion of the GB}
\begin{figure}[bht]
   \centering
    \includegraphics[scale = 0.05]{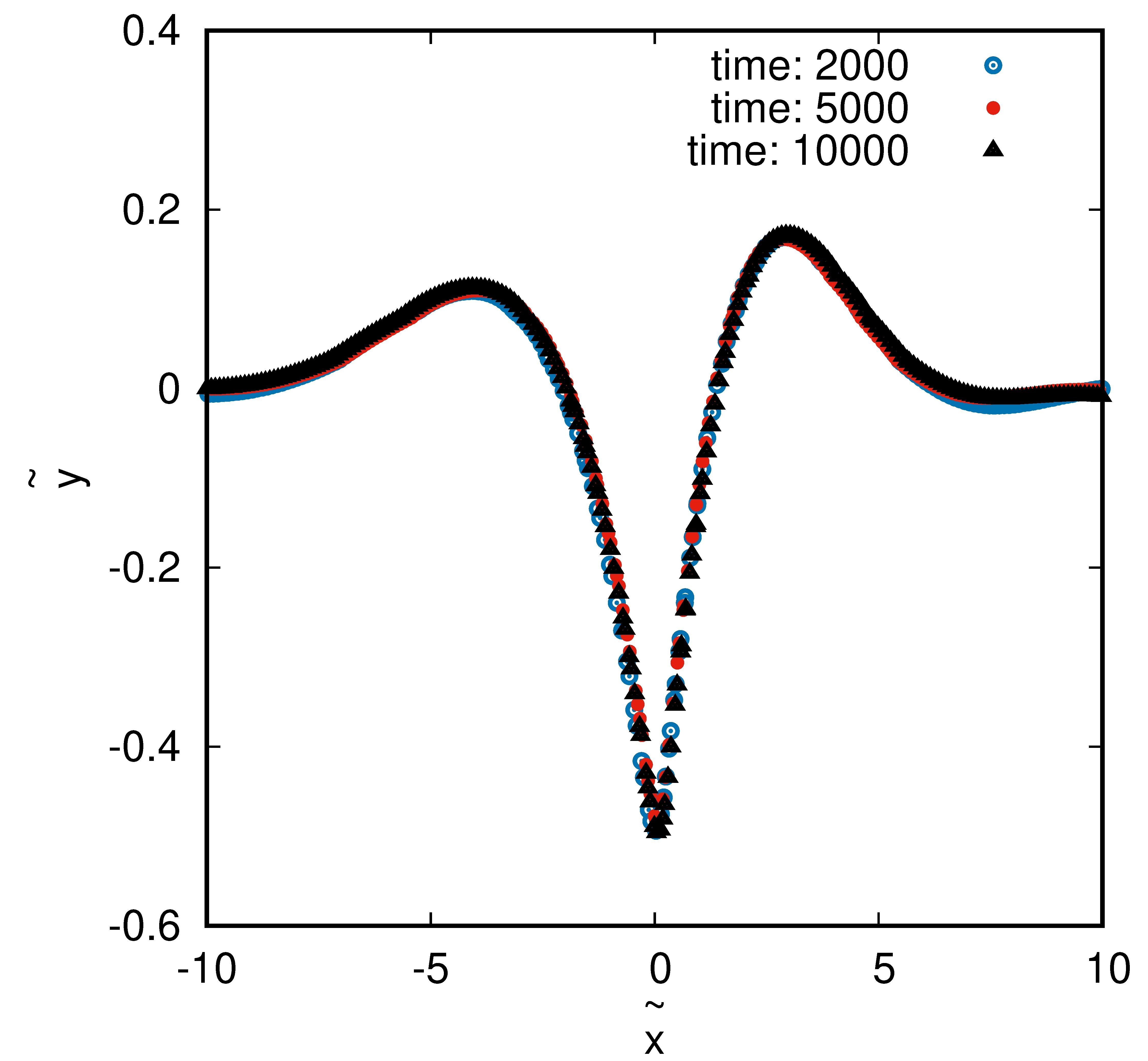}
   \caption{Normalized surface profiles at time 2000, 5000, 10000 for applied field $H_m = 3.0\times 10^8 A/m$ where we observe decelerating behavior. Here, the non-steady motion of the boundary also produces self-similar
surface profiles. 
     }
  \label{fig:surfacedecel}
\end{figure}

We have already mentioned in Section~\ref{res2d} that applying the different external magnetic fields results in 
different behavior of the mobile groove, viz. stationary, steady state, and non-steady state. We have also noticed 
that an applied magnetic field of magnitude $H_m = 3.0\times 10^8 A/m$ generates a non-steady state-type groove 
behavior. Section~\ref{res2d} mainly focuses on the steady-state type behavior of the mobile groove. We 
observe the corresponding surface profiles are self-similar, and the profiles match Mullins' analytical 
solution. In this section, we discuss the non-steady state behavior. Fig.~\ref{fig:surfacedecel} shows the  
normalized surface profiles at times 2000, 5000, and 10000 when we apply an external magnetic field of $H_m = 
3.0\times 10^8 A/m$. Unlike the asymmetric profiles observed during the steady-state grain boundary motion where 
there is a minimal
accumulation of solute on one side (left) (Fig.~\ref{free32}), solute accumulates on
both sides of a decelerating boundary, forming
hills on either side. 
These profiles also significantly deviate and display complete different features from those of a symmetric stationary groove. It is also evident that, for all
profiles, the hills in the direction of boundary motion (right side of
the boundary) are more prominent. 
Furthermore, the surface profiles for the non-steady state mobile groove also possess self-similarity. 
Experimental investigations on grain boundary migration in Mo and NiAl polycrystals report
similar profiles~\cite{rabkin2004scanning,gladstone2001grain}.

\section{\label{sec:sup2} Energy difference between the grains in the 3D system}
\begin{figure}[htbp]
    \centering
    \subfloat[]{\label{ener1}\includegraphics[width=0.4\linewidth]{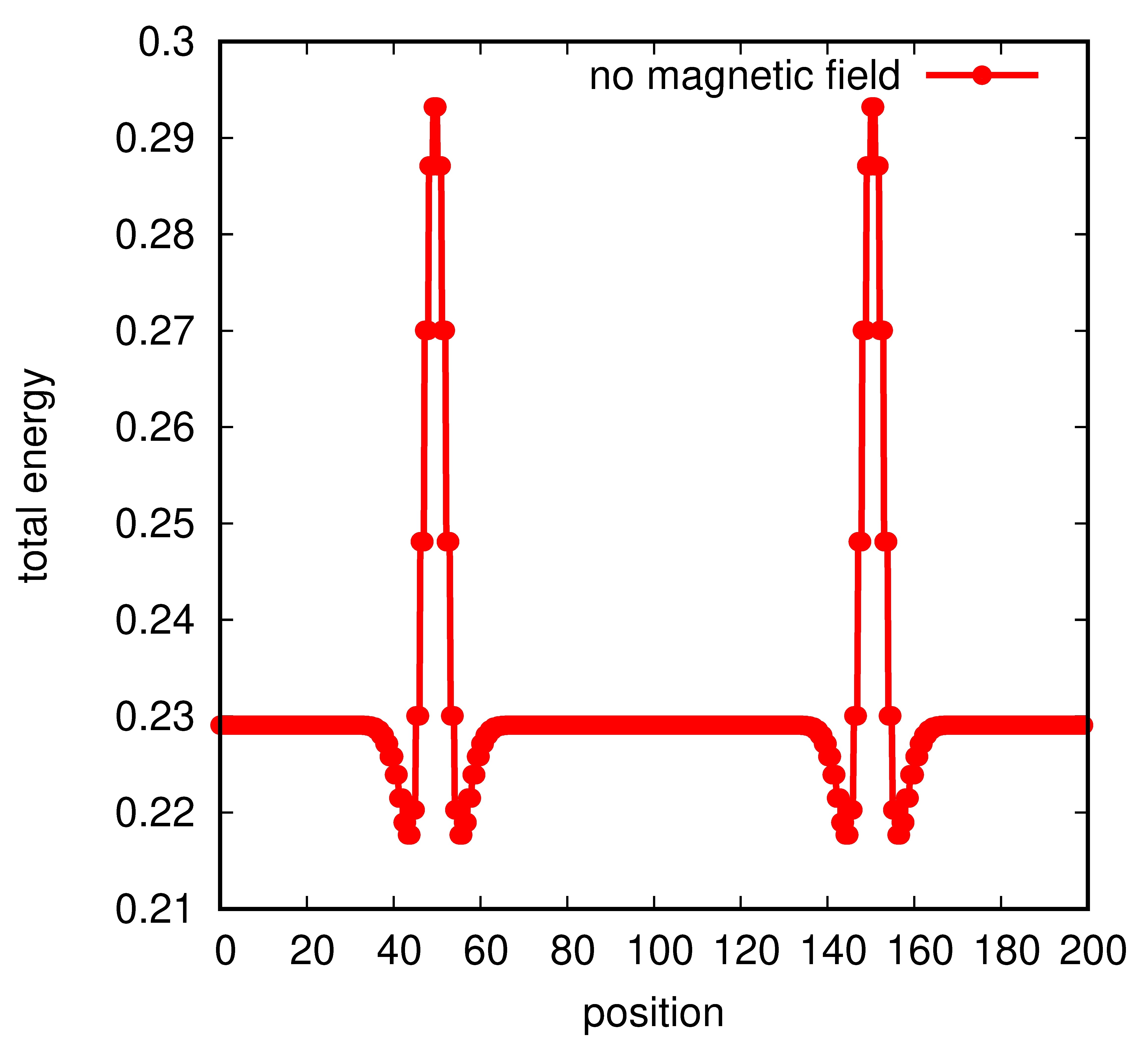}}\hfill
     \subfloat[]{\label{ener3}\includegraphics[width=0.4\linewidth]{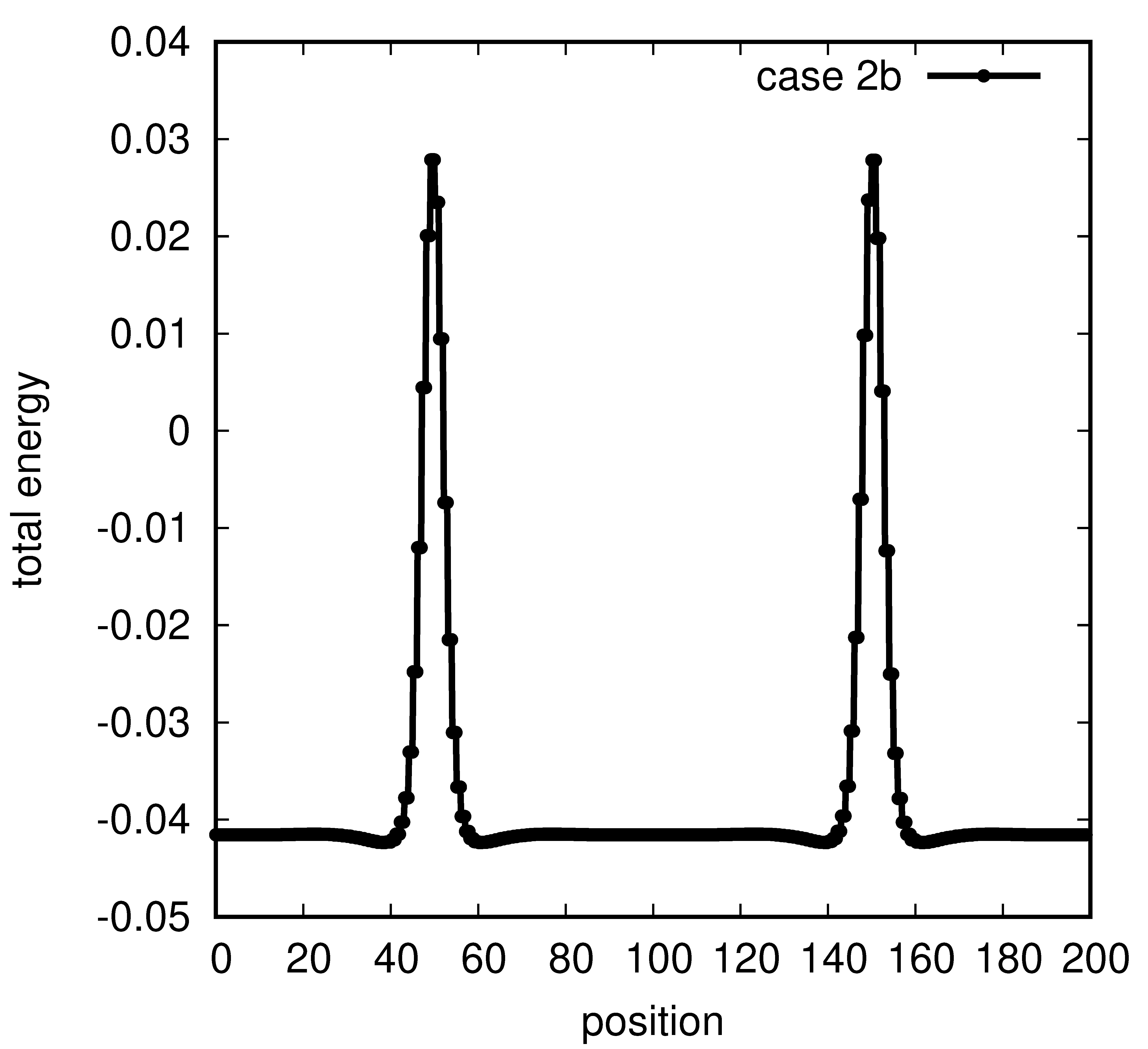}}\hfill
\subfloat[]{\label{ener2}\includegraphics[width=.4\linewidth]{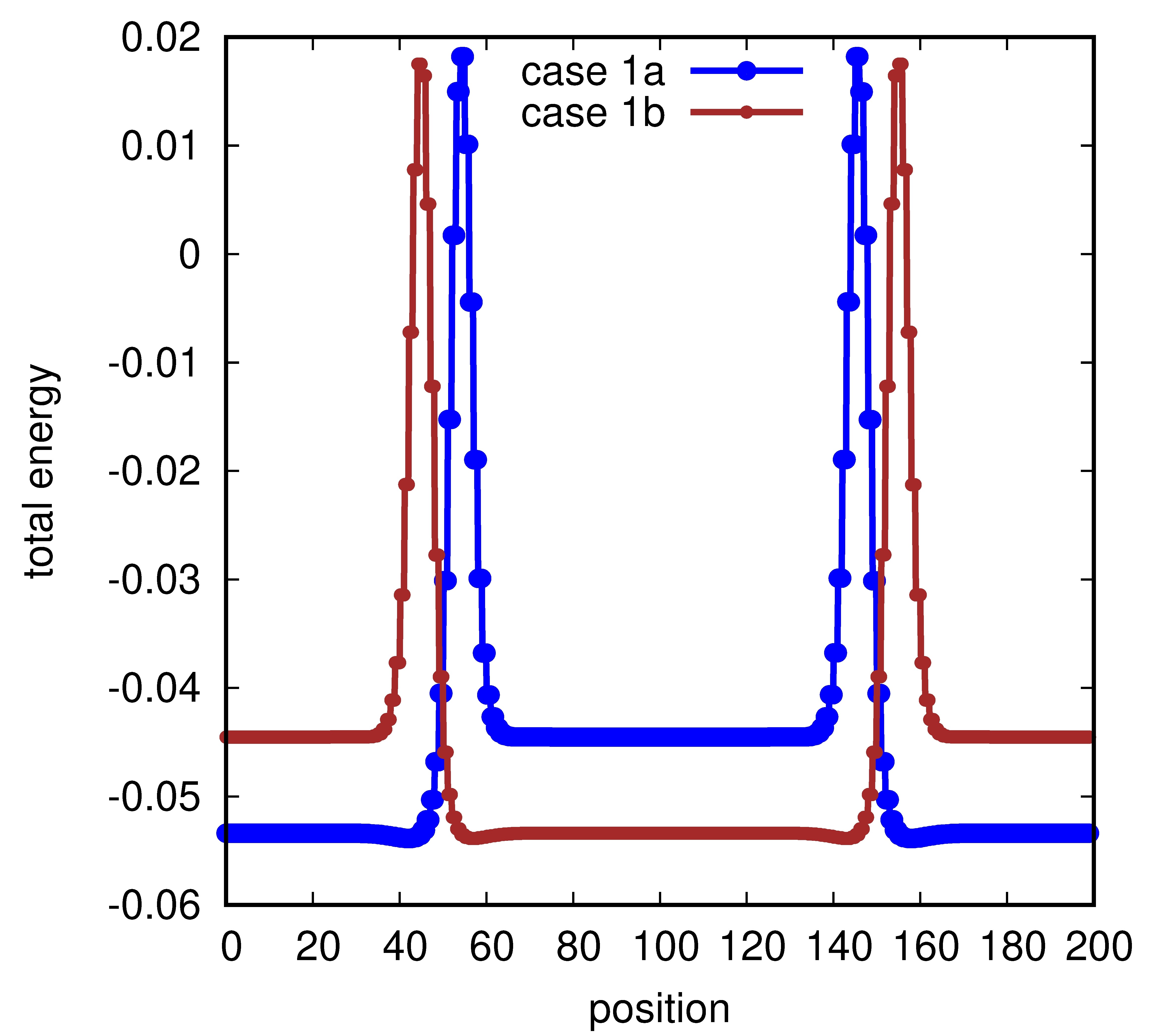}}\hfill 
    \caption{Energy difference between different grains for (a) the system without applied magnetic field (red), (b)case 2b (black), (c) case 1a (blue), and case 1b (brown), respectively. The line is plotted along $x$ direction at $ny/2$ and $nz/2$ in the case of 3D hexagonal polycrystal system.}
    \label{fig:ener}
\end{figure}

Fig.~\ref{fig:ener} shows the energy difference between the grains of the hexagonal system. Fig.~\ref{ener1} compares the energy between the grains for the case with no applied magnetic field and Fig.~\ref{ener3} for case2b. 
In Section~\ref{3d}, we have discussed that for case 2b, the system behaves similarly to the system with no applied magnetic field. From Fig.~\ref{ener1} and Fig.~\ref{ener3}, it is evident that the energy difference between both grains does not change for both cases. However, the corresponding total energy for case 2b is lower (due to a magnetic field) than the system without a magnetic field.
Since the growth and shrinkage of a specific grain are governed by the difference in the energies among the corresponding grains, the grains neither grow nor shrink in these cases.

On the other hand, Fig.~\ref{ener2} demonstrates the cases where Grain 4 either grows (case 1b) or shrinks (case 1a). It can be explained by considering the energy difference between the grains. From Fig.~\ref{ener2}, it is clearly 
understood that for case 1a, Grain 4 possesses higher energy than the neighboring grains (marked by blue); thus, it shrinks. On the contrary, it possesses lower energy for case 1b (marked by brown), which triggers the growth of Grain 4 at the expense of the others.

\newpage
\bibliographystyle{unsrt}
\bibliography{scopus}% Produces the bibliography via BibTeX.